\documentclass{aa}  
\usepackage{graphicx}
\usepackage{psfig}
\usepackage{amsmath}
\usepackage{aas_macros}
\usepackage[authoryear]{natbib}
\def\e20{$\times 10^{20}$}
\def\ergs{erg~s$^{-1}$}

\def\today{\number\day -\number\month -\number\year}

\def\hi{H\,{\sc i}}
\def\eso{ESO~2400100}
\def\n70{NGC~7070A}

\begin{document}
   \title{XMM--{\it Newton} X-ray and Optical Monitor far UV observations of 
   \\ NGC~7070A and ESO~2400100 shell galaxies\thanks{Based on XMM--{\it Newton} 
   observations}}
\titlerunning{X-ray and far UV observations of shell galaxies}
\authorrunning{Trinchieri et al.}

   \author{G. Trinchieri\inst{1}, R. Rampazzo\inst{2}, C. Chiosi\inst{3},  R. Gr\" utzbauch\inst{4},  
   A. Marino\inst{2} 
         \and   R. Tantalo\inst{3}
          }

   \offprints{R. Rampazzo}

   \institute{INAF-Osservatorio Astronomico di Brera,
              Via Brera 28, I-20121, Milano, Italy \\
             \email{ginevra.trinchieri@brera.inaf.it} 
       \and
              INAF-Osservatorio Astronomico di Padova,
              Vicolo dell'Osservatorio 5, I-35122, Padova, Italy \\
              \email{roberto.rampazzo@oapd.inaf.it, antonietta.marino@oapd.inaf.it}
       \and         
              Dipartimento di Astronomia Universit\`a di Padova, 
            Vicolo dell'Osservatorio 2, 35122 Padova, Italy\\
 \email{cesare.chiosi@unipd.it; rosaria.tantalo@unipd.it}
         \and
           Institute for Astronomy (IfA), University of Vienna,
              T\"urkenschanzstrasse 17, A-1180 Vienna, Austria\\
              \email{gruetzbauch@astro.univie.ac.at} 
             }

   \date{Received, March 2008; accepted, 2008}

 
  \abstract
 {Shell galaxies are considered the debris of recent accretion/merging 
   episodes. Their high frequency in low density environments suggests 
   that such episodes could drive the secular evolution for at least some 
   fraction of the early-type galaxy population.}
   {We present  XMM-Newton X-ray  observations of two shell galaxies, 
   NGC~7070A and ESO~2400100, and far UV observations obtained with the Optical
   Monitor for these and for an additional shell galaxy, NGC 474, for which we 
   also have near and far UV data from GALEX.  We aim at
   gaining insight on the overall evolution traced by their star formation 
   history and by their hot gas content.} 
   {The X-ray and the far UV data are used 
   to derive  their X-ray spatial and spectral characteristics and 
   their UV  luminosity profiles.  We use models developed ad hoc to investigate
   the age of the last episode of star formation from the (UV - optical) colors
   and line strength indices. 
   }
   {The X-ray spatial and spectral analysis show significant differences in the 
   two objects. 
   A low luminosity nuclear source is the dominant component in NGC~7070A 
   (log L$_X$=41.7 erg~s$^{-1}$ in the 2-10 keV band).  In ESO~2400100,
   the X-ray emission is due to a low temperature plasma with a contribution 
   from the  collective emission of individual sources. 
   In the Optical Monitor image 
   ESO~2400100 shows a double nucleus, one bluer than the
   other.  
This probably results from  a very recent star formation event in the northern nuclear region.
   The extension of the UV emission is consistent with the
   optical extent for all galaxies, 
   at different degrees of significance in different
   filters. 
The presence of the double nucleus, corroborated by  the (UV - optical) 
colors  and line strength indices analysis, suggests that ESO~2400100 is 
accreting a faint companion.  We explore the evolution of the X-ray luminosity 
during accretion processes with time. We discuss the link between the presence 
of gas and age, since gas is detected either before coalescence or 
several Gyr ($>3$) after. 
}
{}

  \maketitle
%

\section{Introduction}

In $\Lambda$CDM cosmology, galaxies are assembled hierarchically over an
extended period by mergers of smaller systems. In this framework, early-type
galaxies showing {\it fine structure} occupy a special position since
they are a testimony of the effects of past merging/accretion events,
and as such they fill the gap between on-going mergers and the relaxed
elliptical galaxy population.

Among examples of {\it fine structure}, 
shells are faint, sharp-edged stellar features \citep{Malin83}
that characterize a significant fraction ($\approx$ 16.5\%) of the 
field early--type galaxy
population \citep{Malin83, Schweizer92,Reduzzi96,Colbert01}. 

Different scenarios for the origin of shells 
emerge from the rich harvest of
simulations performed since their discovery in the early 80's, mostly 
involving galaxy-galaxy interactions.  These range  from 
merging/accretion  between galaxies of different morphological type or masses
 \citep[mass ratios typically 1/10 - 1/100, ][]{Quinn, Dupraz86,
Hernquist87a,Hernquist87b} to significantly weaker interaction events
\citep{Thomson90,Thomson91}.  A few models that invoke gas ejection due to the  central AGN or the power of supernovae \citep{Fabian80, Williams85} do not associate the shell formation with environment.  Most
models can reproduce qualitatively basic characteristics such as spatial
distribution, frequency and shape of observed shell systems \citep[see
e.g.][and references therein]{Wilkinson00,Pierfederici04,Sikkema07}.
In the accretion models,
the more credited scenario for their formation,  shells are density
waves formed  by  infall of stars from a companion.  
A major merger may also  produce shells
\citep{Barnes92,Hernquist92,Hernquist95}.  The fact that
shells are frequently found in the field, rather then in the cluster
environment, finds a direct explanation in the week interaction
hypothesis, since the  galaxy-galaxy ``harassment'' within the 
cluster tends to destroy shells, while poorer environments are much better suited, because the group velocity dispersion is
of the order of the internal velocity of the member galaxies
\citep[see e.g.][]{Aarseth, Barnes85, Merritt}.
The class of the so-called ``internal models'' for shell formation  suggests that star formation within a giant shell is the result of shocked interstellar gas.
In such a case shells are expected to be bluer than the parent galaxy,
up to $\approx$0.5 mag in U-B and B-V \citep[see e.g.][]{Fabian80}.
In a few cases this has been observed,  
\citep[see e.g.][and reference therein]{Sikkema07} but attributed to
multiple accretion events of galaxies of different intrinsic color.

In the framework of a hierarchical cosmology, all galaxies, including
early-type galaxies, are expected to contain multi-epoch stellar
populations.  Shell galaxies, among the early-type class of  
galaxies, are the ideal candidate
to contain also a young stellar population since some simulations  
indicate a shell dynamical age of 0.5--2 Gyr \citep{Hernquist87a,Nulsen89}.
The star formation history of shell early-type galaxies has
been analyzed by \citet{Longhetti00} using line--strength indices. They
show that shell-galaxies encompass the whole range of ages inferred
from the H$\beta$ vs. MgFe plane, indicating that among them recent and
old interaction/acquisition events are equally probable. If shells are
formed at the same time at which the ``rejuvenating'' event took place,
shells ought to be long--lasting phenomena. Recently, \citet{Rampazzo07}
combining {\it GALEX} far UV data and line--strength indices  show
that the peculiar position of some shell galaxies in the (FUV-NUV)
vs. H$\beta$ plane could be explained in terms of a recent (1-2 Gyr old)
rejuvenation episode.

A rejuvenation of stellar population requires the presence of fresh
gas during the accretion event. Therefore, a study of the cold, warm and hot gas 
phases is important in order to consider many of the 
elements relevant  for the evaluation of shell galaxy evolution.
\citet{Rampazzo03} found that the warm ionized (H$\alpha$)
gas  and stars appear often decoupled suggesting an external acquisition 
of the gas, as predicted by merging models \citep{Weil93}.
At the same time, a set of observations showing a clear association
between cold (\hi/CO) gas and stars challenge present merging models
which do not predict it unless cold gas behaves differently from the
ionized gas \citep{Schiminovich94,Schiminovich95,Charmandaris00,Balcells01}.

The bulk of the interstellar medium (ISM)
in early-type galaxies emits in X-rays,
 and only comparatively small quantities are
detected in the warm and cold phases of the ISM \citep{Bregman92}. 
The hot ISM  is believed to build up primarily through stellar mass loss ($\approx$ 1 M$_\odot$/yr  in an old passively evolving elliptical 
galaxy with M$_B=-22$~mag) and 
Type Ia supernova ejecta  \citep[$\approx$0.03 M$_\odot$/yr in a galaxy with M$_B=-22$~mag; see details in][]{Greggio05}.
The two processes produce an almost identical mass in metals 
although the SNIa ejecta are dominated by Fe.
\citet{HB06}, using proper
spectral fits to excellent Chandra data,  were able to better determine  the  
metal abundances of the X-ray emitting gas for a sample of early-type galaxies. 
By estimating  stellar abundances from  optical line strength indices, adopting 
simple stellar population models, they showed that the hot ISM and the stars 
have similar abundances.  

The link between the stellar and ISM metallicities
could be masked/contaminated by the accretion  of a gas-rich system.
At the same time, the content of hot X-ray emitting gas could be  correlated with the ``age'' of the rejuvenation episode. Early-type galaxies
with {\it fine structure} (e.g. shells), which are considered
{\it bona fide} signatures of the ``dynamical youth'' of the galaxy,
tend to be less X-ray luminous than more relaxed, ``mature'' ellipticals with
little evidence of  {\it fine structure}. These are also characterized
by extended, and generally stronger, X-ray emission \citep{Sansom00}.
Simulations suggest that mergers \citep[see e.g.][]{Cox06} or interactions
\citep[see e.g.][]{dercole00} among galaxies produce indeed a wide range of
X-ray luminosities. One extreme example in this picture is  NGC 474, which 
shows  a well developed shell system and has an X-ray luminosity
consistent with the low end of the expected emission from discrete sources
\citep{Rampazzo06}. The X-ray domain may further disclose an otherwise
hidden AGN activity, that could also be a result of the merging episodes.
Again an extreme example of nuclear activity in a merger remnant
can be found in the newly discovered spectacular shell system
in the elliptical host of the QSO MC2 1635+119 by \citet{Canalizo07}.

So, whereas accretion/merging events are widely believed to be at
the origin of shell galaxies, all the details such as: the age of the
event and duration of the shell structure, the global secular evolution
of the stellar and gas components of the host galaxy as well as the timing
for triggering the AGN, its duty cycle and feedback, are far
from being firmly established.  In light of the high fraction of
shell galaxies in the field, interaction/accretion/merging events seem to have
played a significant role in the evolution of the early-type class as
a whole. At the same time, there is still  the general open  question 
of whether a link exists between shell galaxies and the early phases 
of merging processes (ULIRGs, AGN, E+A galaxies etc.) on one side 
and the general class of ``normal'' early-type  galaxies on the other.

In the above framework, we discuss the X-ray (XMM-{\it Newton}) observations 
of two shell galaxies, NGC~7070A and ESO~2400100, 
taken from the \citet{Malin83} compilation. We further present far UV 
XMM-{\it Newton}  Optical Monitor (OM) observations of these objects, and we
add {\it GALEX} far UV observations of NGC~474, 
for which the results from XMM-Newton X-ray observations 
have been already presented in  \citet{Rampazzo06}.
Through far UV photometry we aim at inferring whether these galaxies
have ongoing/recent star formation activity and its distribution 
across the galaxy. We finally aim to correlate the above information 
 with those extracted from their hot gas content and properties in 
light of our current understanding of these components.

The plan of the paper is  the following. Section~2 describes the
relevant properties of our sample gathered from the 
literature. Section~3  presents the X-ray and Far UV observations and 
data reduction. Results are presented in Section~4 and discussed in
Section~5. H$_0$=75 km~s$^{-1}$ is used throughout the paper.

\begin{table*}
\caption{Relevant photometric, structural and kinematic  properties}
\begin{tabular}{llllc}
\hline\hline
&  NGC~474     & NGC~7070A &ESO 2400100   & Ref. \\
\hline
Morphol. Type          &  (R')SA(s)0    &(R')SB(l)0/a & SAB0: pec  & [1] \\
Mean Hel. Sys. Vel. [km~s$^{-1}$] &2366$\pm$16 &2391$\pm$18 &3184$\pm$14 & [2] \\
Adopted distance [Mpc]         &  32.5           & 31.9 & 42.4 & [3]\\
$\rho_{(x,y,z)}$ [gal Mpc$^{-3}$] & 0.19            & & & [3] \\
Environment                             & Cetus-Aries  & & & [3] \\
cloud                                   &  52-12          & & & [3]\\
                                         &                    & & &\\
{\bf Apparent magnitude,}          &                    & & &\\
{\bf colours, indices}:            &                    & & &\\
B$_T$                            & 12.36$\pm$0.16& 13.35$\pm$0.15&12.63$\pm$0.24 & [2] \\
K$_T$ (2MASS)         & 8.555$\pm$0.039 &9.130$\pm$0.026 & 8.698$\pm$0.028 & [1] \\
6 cm [mJy]                             &            & 1.3   &         &   [5] \\
S$_{60\mu m}$ (IRAS) [Jy] & & 0.26$\pm$0.040 & & [1]\\
S$_{100\mu m}$ (IRAS) [Jy] & &0.75$\pm$0.077 & & [1]\\
Mg2                           &            &         & 0.28/0.21 & [6]\\
H$\beta$                   &            &         & 1.54/2.79 & [6]\\
$\Delta4000$                &            &         & 2.31/2.01 & [6]\\
H+K(CaII)                 &            &         & 1.18/1.24 & [6]\\
H$\delta$/FeI              &            &         & 0.97/0.88 & [6]\\
                              &                  & & & \\
{\bf Galaxy structure}:                  &                  & & & \\
Effective Surf. Bright. $\mu_e$(B) &21.99$\pm$0.31&22.48$\pm$0.33 &21.84$\pm$0.33 &[2]\\
Diam. Eff. Apert., A$_e$ [\arcsec] &  64.6      & & &[2] \\
Average ellipticity      & 0.21    &0.28 & 0.46 & [2]\\
P.A. [deg]                  & 75   &6.5 & 132.5& [2]\\
Fine structure ($\Sigma$) & 5.26     & & & [4]\\ 
                                 &                           &  & &   \\
{\bf Kinematic parameters}        &                        & & & \\
Vel.disp. $\sigma_0$ stars [km~s$^{-1}$]  &163.9$\pm$5.1& 101.0$\pm$20.2 & 225/223&[1,6] \\
App.Max. rotation V$_{max}$ star [km~s$^{-1}$] & 30$\pm$6 &0.0$\pm$0.0 & & [2] \\
App.Max. rotation V$_{max}$ gas [km~s$^{-1}$] &158.4$\pm$9.3 & & & [2] \\
                          &                           & & & \\
\hline
\end{tabular}
\label{table1}

\medskip
References: [1] {\tt NED http://nedwww.ipac.caltech.edu/};
[2] {\tt HYPERLEDA http://leda.univ-lyon1.fr/};  
[3] \citet{Tully88} (H$_0$=75 km~s$^{-1}$~Mpc$^{-1}$);
[4] \citet{Sansom00}; [5] \citet{Sadler89}; [6] \citep{Longhetti99,Longhetti00} data refer
respectively to the $a/b$ nuclei embedded in the envelope of ESO~2400100.
\end{table*}
 
\section{The sample}

In the following paragraphs and in Table~\ref{table1} we present the
main characteristics of the galaxies considered by the present study
derived from the recent literature.

\underbar{NGC 7070A}~~~~ 
This lenticular galaxy  at a heliocentric
velocity of 2450$\pm$20 km~s$^{-1}$ has a bright Scd companion,
NGC7070, at a projected separation of 21\arcmin\  (195 kpc) 
and at least two other smaller spirals near NGC7070 \citep{Oosterloo02}. 
\citet{Malin83} describe its shells as nearly circular  and 
about complete. The shells are probably the remnants of the same event 
which formed the strong dust--lane which crosses the entire galaxy.  
\citet{Brosh85} attribute the dust--lane to the
 accretion, about 1~Gyr ago, of a small disk galaxy containing 
at least $\sim$ 4 $\times$ 10$^7$ M$_{\odot}$ of gas and dust. 
\citet{Sharples83} performed an
analysis of its kinematics along several axes, including apparent 
major and minor axes.
The stellar component shows no rotation at any position angle
(V$_{max} <$ 30 km~s$^{-1}$). The velocity dispersion is very low,
$\sigma \approx$ 100 km~s$^{-1}$, and does not change with radius. In
view of the above kinematical properties, \citet{Sharples83} 
suggest that NGC7070A may be a face-on disc galaxy with a 
triaxial or prolate bulge and that the dust lane is unlikely to
be in an equilibrium configuration. \citet{Sharples83} 
measured also the gas kinematics along the
dust-lane showing that the rotation curve is approximately that of a
solid body. Since there is no evidence of rotation of the stellar
component, they conclude that the angular momentum axes of 
the stars and gas must be very different.
\citet{Rampazzo03} showed indeed that the ionized gas distribution in NGC~7070A 
is elongated and has an asymmetric structure relative to the stellar body.
The nucleus of NGC~7070A is a faint radio emitter at 4.885 GHz  
\citep{Sadler89}. No \hi\ is detected down to the level of M$\sim 
5 \times 10^7$ M$_{\sun}$ \citep[][using the
Australia Telescope Compact Array]{Oosterloo02}. 

\begin{figure*}
\centering
\includegraphics[width=16cm]{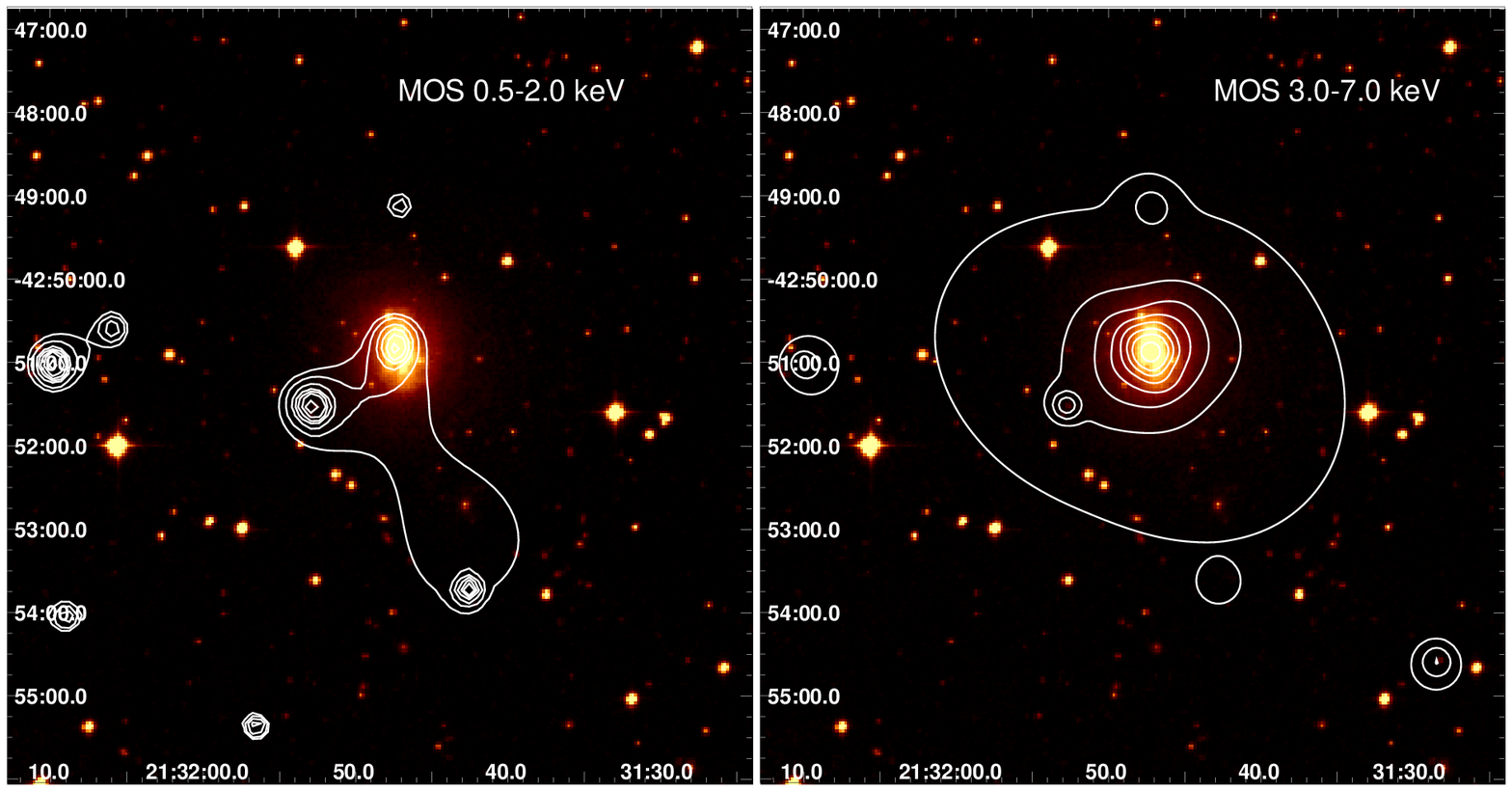}
\includegraphics[width=16cm]{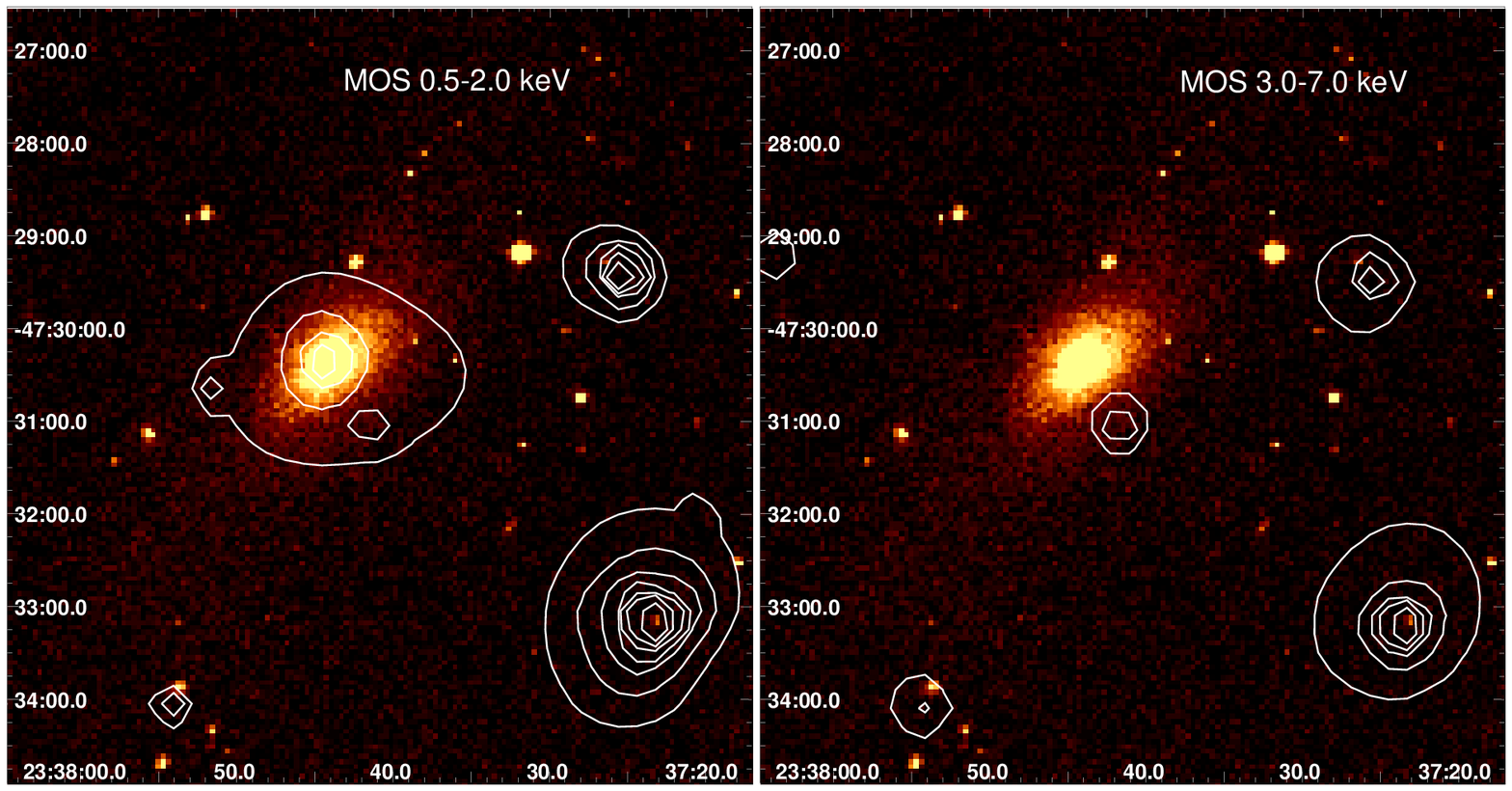}
\caption{Isointensity contours from the adaptively smoothed XMM-{\it
Newton} combined MOS image in two broad energy bands, hard and soft, on
the DSS2 plate for NGC~7070A ({\it top panels}) and ESO2400100 ({\it
bottom panels}).}
\label{fig1}
\end{figure*}

 \underbar{ESO 2400100}~~~~\citet{Longhetti98b} found two 
 distinct  nuclei separated by $\approx 5''$  and with systemic 
 velocities different by 200 km s$^{-1}$.  \citet{Rampazzo03} show that 
 the ionized gas  and the stars are completely decoupled. 
Simulation of inter-penetrating  encounters  by \citet{Combes95} and 
 references therein  show that the U-shape and the correlated inverted 
 U-shape velocity profile along the line connecting the two galaxy nuclei 
 is a bona fide signature of the ongoing interaction.  Since
the gas 2D morphology, the stellar and the gas kinematics
suggest that the two nuclei are still interacting, if
the development of the shells is connected with these interactions, their
formation process is still on-going.

\underbar{NGC~474}~~~~\citet{Rampazzo06} have already presented this
system in details, along with its X-ray properties. NGC~474,
also known as Arp 227, is interacting with NGC~470, a spiral galaxy at
$\approx$ 5.4\arcmin\ to the west. 
The two galaxies have a small (virtually null) velocity
difference.  The surface photometry of \citet{Shombert87} indicates that
NGC~474 is a S0 galaxy with a smooth and undistorted inner luminosity
profile. \citet{Turnbull99} suggests that the colors of the inner shells
follow that of the galaxy while the outer shells appear to
be bluer.  Similar results are found by \citet{Pierfederici04} and
\citet{Sikkema07}.  Using ACS-HST, these latter authors suggest
that the morphology of the shell system is probably due to multiple
accretion events.  NGC~474 is a very gas-poor system. It  has not been
detected in H$\alpha$ and also the hot gas component is poor: it's X-ray
emission is very low, and its luminosity, L$_x \sim 10^{39}$ \ergs, 
is consistent with the low end of the expected emission from
discrete sources \citep{Rampazzo06}.

\begin{figure*}
\resizebox{18cm}{!}{
 \includegraphics{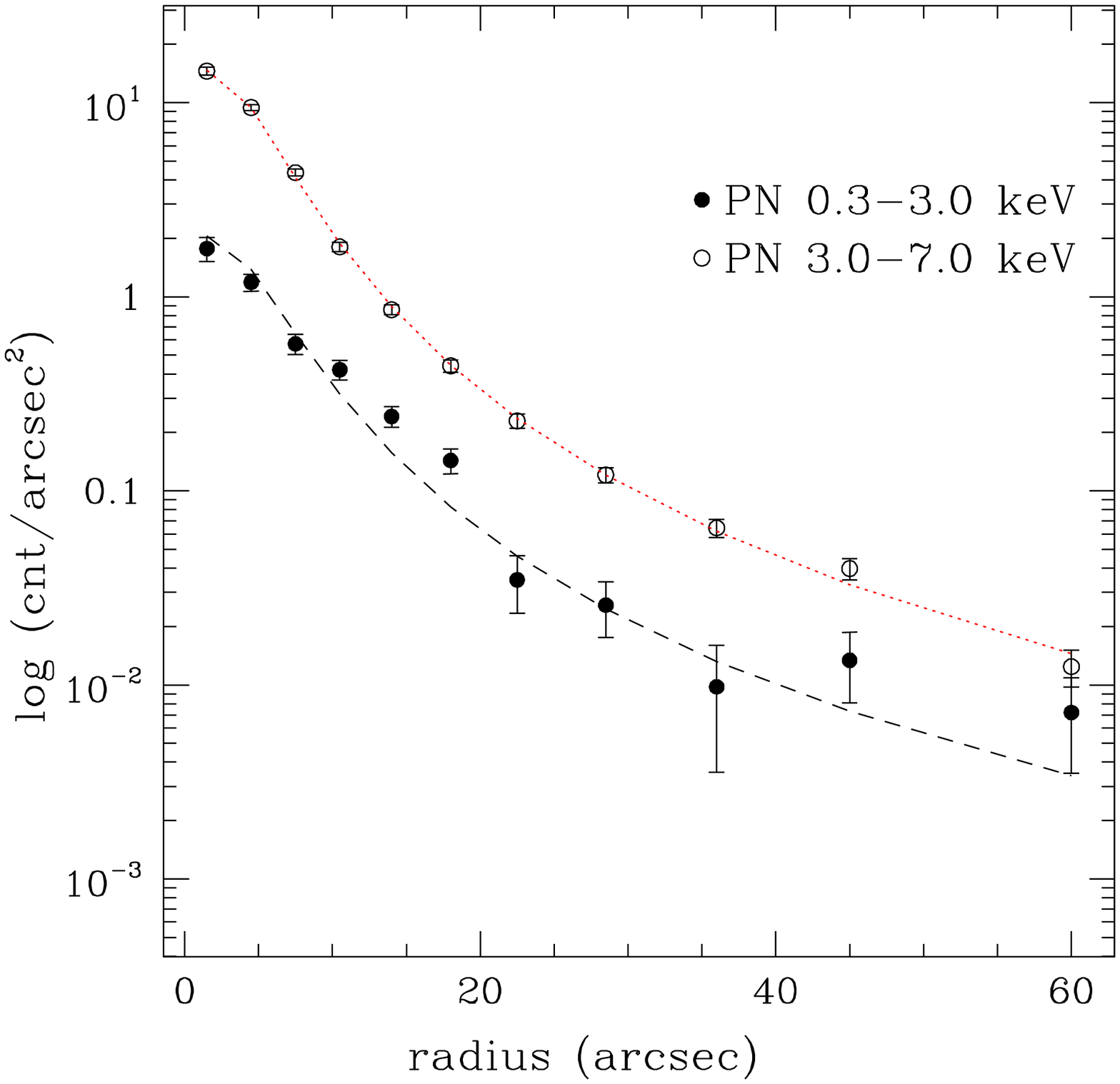}\\
 \includegraphics{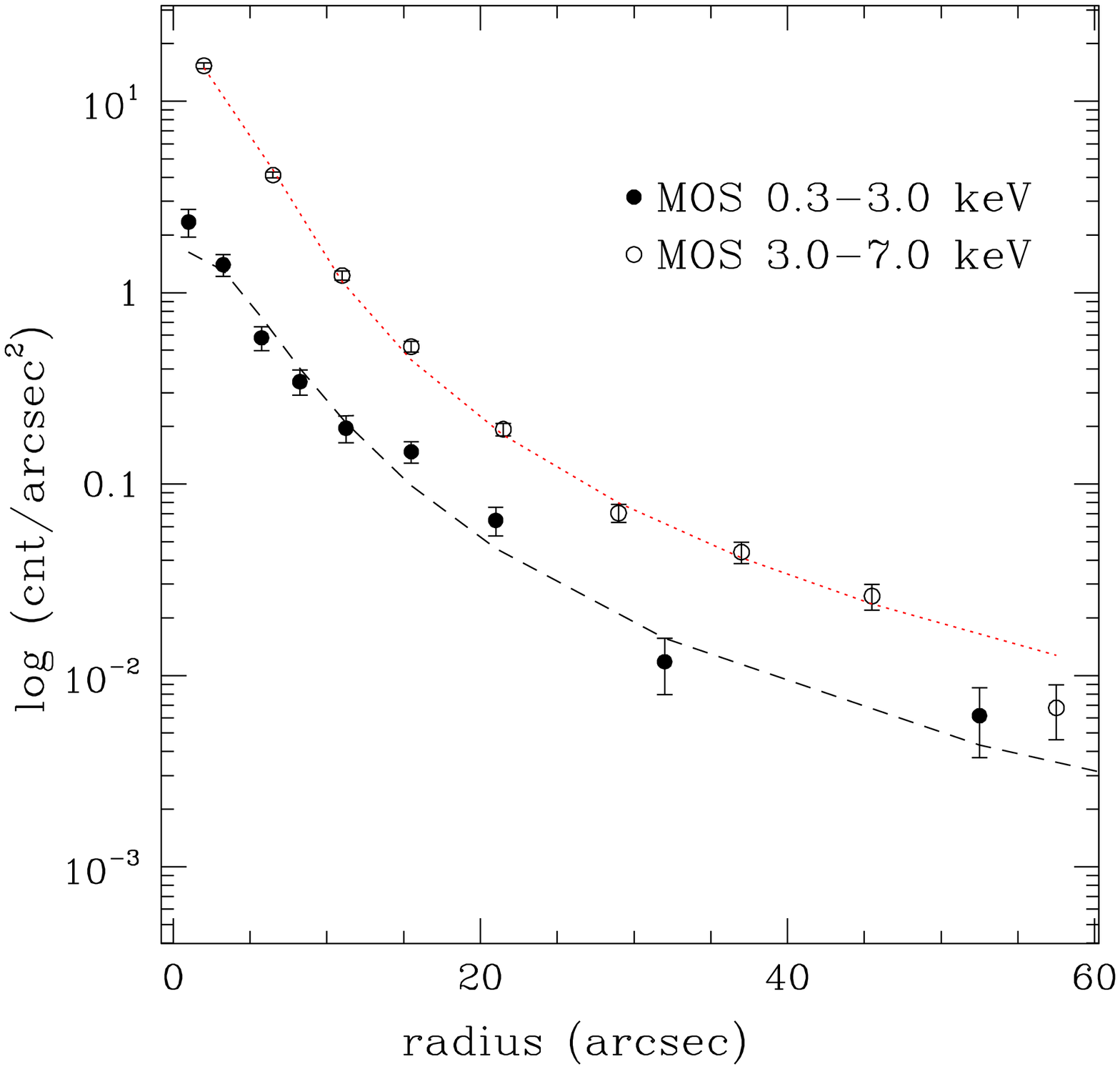}}
\caption{Azimuthally averaged radial profile of the net emission in NGC7070A
in two energy bands from the EPIC-pn and the EPIC-MOS combined data.  
The background is estimated in an annulus outside the outermost radius considered. 
The dashed and dotted lines correspond to the parametrization of the
profiles with a $\beta$-model with r$_0 \sim 5'' $ and $\beta \sim
$0.6, consistent with the EPIC response. 
}
\label{fig2}
\end{figure*}

Beside the shell structure, spectacular in NGC 474, these systems have
other characteristics in common.  They are all S0 galaxies located in
low density environments.  They are all interacting systems, although at
different levels of interaction: NGC~7070 and NGC~7070A are well 
separated and mildly interacting, while the interaction is strong
between NGC~470 and NGC~474 as shown by the \hi\ data.  
Two nuclei are embedded and interacting within a single envelope in  
ESO~2400100. Current literature supports the 
accretion hypothesis for all galaxies.  There is strong evidence that the current interaction is not responsible for the origin of the shell system in 
either NGC~474 or NGC~7070A,   at variance with ESO~2400100. 
\citet{Pierfederici04} have shown that the radial distribution of 
the shell surface  brightness in NGC 474 is inconsistent with the weak interaction hypothesis.  There is also kinematical evidence that the 
strong dust--lane which crosses the entire body of NGC~7070A,
 with an orientation intermediate between the optical
major and minor axes, is the result of external
accretion \citep{Sharples83}, which could be related with the formation of the
shell system.
At the same time, our galaxies are not homogeneous in
mass\footnote{The velocity dispersion, $\sigma$, does not have a
one-to-one relationship with the mass of the galaxy. Recent work by
\citet{Cappellari06} has investigated the relation between the galaxy
mass-to-light (M/L) ratio and the line-of-sight component of the
velocity dispersion with the effective radius. They provide relations
(their equations 7 and 10) which allow a transformation from measured
velocity dispersion to galaxy mass; \begin{equation} M_{10} = (16.5 \pm
7.8)\;\sigma_{200}^{3.11 \pm 0.43} \end{equation} where $M_{10}$ is the
galaxy mass in units of $10^{10}\,\rm M_{\odot}$ and $\sigma_{200}$ is
the velocity dispersion in units of $200 \;\rm km\,s^{-1}$. The errors
have been propagated from the parameter values found by
\citet{Cappellari06}.}: their stellar velocity dispersion is in the
range 101 $ < \sigma  <$ 225 km~s$^{-1}$  (see Table~1), corresponding
to masses of 0.2 $<M< $ 2.4 $\times$10$^{11}$ M$_\odot$.

\begin{table*}
\tiny
\caption{Journal of the XMM-Newton EPIC and  Optical Monitor
observations. All were part of the same accepted proposal (ID: 0200780,
PI: Trinchieri).  The X-ray data for NGC 474 have already been presented
and discussed in \citet{Rampazzo06}. Exposure times for the EPIC
instruments refer to the clean dataset, after removal of background
flares.}
\begin{tabular}{llllllllllll}
 \hline
 \multicolumn{1}{l}{Galaxy}& 
 \multicolumn{1}{c}{EPIC}  &
 \multicolumn{1}{c}{EPIC}  &
\multicolumn{1}{c}{OM}  &
\multicolumn{1}{c}{OM}  & 
\multicolumn{1}{c}{OM}  &
\multicolumn{1}{c}{OM}  &
\multicolumn{1}{l}{observing} &
\multicolumn{2}{c}{GALEX}  &
\multicolumn{1}{l}{observing} \\
 \multicolumn{1}{l}{}& 
 \multicolumn{1}{c}{PN}  &
 \multicolumn{1}{c}{MOS}  &
\multicolumn{1}{c}{UVM2}  &
\multicolumn{1}{c}{UVW1}  & 
\multicolumn{1}{c}{U}  &
\multicolumn{1}{c}{B}  &
\multicolumn{1}{c}{date} &
\multicolumn{1}{l}{FUV} &
\multicolumn{1}{l}{NUV}&
\multicolumn{1}{c}{date} 
\\
\multicolumn{1}{c}{}&
\multicolumn{1}{c}{ exp. [ks]} &
\multicolumn{1}{c}{ exp. [ks]} &
\multicolumn{1}{c}{ exp. [s]} &
\multicolumn{1}{c}{ exp. [s]} &
\multicolumn{1}{c}{ exp. [s]} &
\multicolumn{1}{c}{ exp. [s]} &
\multicolumn{1}{c}{} &
\multicolumn{1}{c}{ exp. [s]} &
\multicolumn{1}{c}{ exp. [s]} &
\multicolumn{1}{c}{}\\
\hline
NGC~474        &   4.3 & 11.4  &   5000 &  5000  &  5000 &  & 2004 -01-24 &
1647 & 1477 & 2003-03-10\\
NGC~7070A    &  26.3 & 30.8  &  4400  &          &         & 4000  &2004-10-28  \\
ESO 2400100  & 20.0  & 26.3  & 4400    &  4400  & 4400 &  &2004-05-11  \\
\hline
\end{tabular}

\label{table2}
\end{table*}

\begin{figure}
\resizebox{9cm}{!}{
\includegraphics{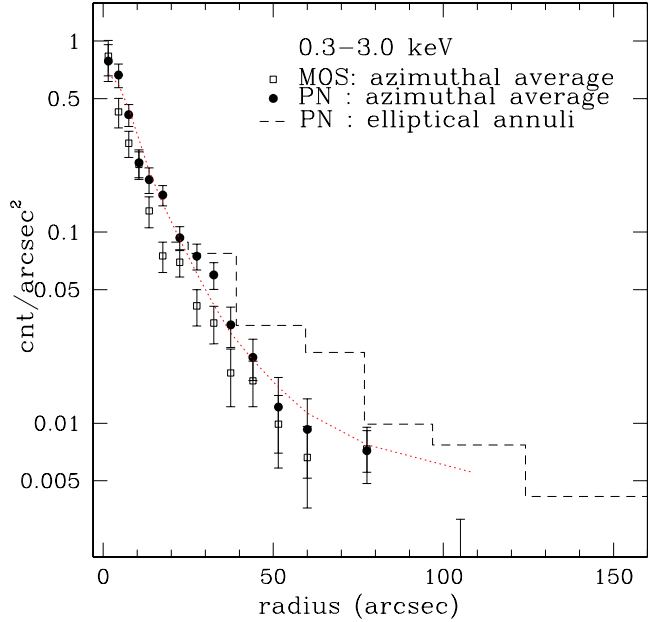}}
\caption{Azimuthally averaged radial profile of the net emission in ESO~2400100
in the 0.3-3.0 keV energy band from the EPIC-pn and the EPIC-MOS combined data. 
The background is estimated in an annulus at $120''-170''$ radii.  
The histogram shows the radial distribution in elliptical annuli aligned to the
elongation visible in the image as a function of the major axis.
The (red) dashed line corresponds to the parametrization of the EPIC-pn
profiles in the assumption of azimuthal symmetry with a
$\beta$-model with r$_0 \sim 6\farcs4 $ and $\beta \sim
$0.5, modulated by the EPIC response, represented by a $\beta$-model with r$_0 =
4\farcs1$ and $\beta=0.63$.
}
\label{fig3} 
\end{figure}
\begin{figure}
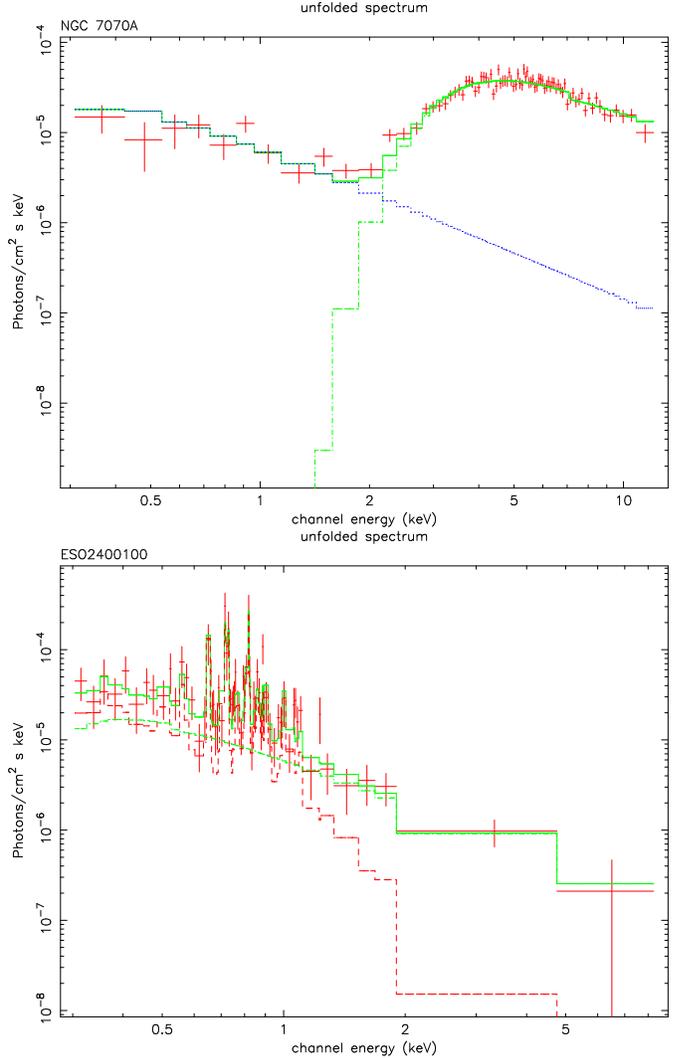

\includegraphics[width=7.0cm,angle=-90]{fig4a.ps}
\includegraphics[width=7.0cm,angle=-90]{fig4b.ps}
\caption{Unfolded spectrum and best fit model (solid histogram). {\it 
Top} NGC 7070A, modeled with an unabsorbed (blue dotted) and a heavily 
absorbed (green dash-dotted) powerlaws with $\Gamma$=1.5 and
N$_H \sim 2 \times 10^{23}$ cm$^{-2}$.  {\it Bottom}
\eso\ modeled by a mekal with kT$\sim$0.5 (red dashed) and 
a powerlaw  with $\Gamma$=1.7 (green dash-dotted), absorbed by N$_H \sim 4
\times 10^{20}$ cm$^{-2}$ above Galactic)
}
\label{fig4}
\end{figure}

\section{Observations and data reduction}

\subsection{X-ray observations}
 
We obtained $\sim30$ ks XMM-Newton observations of NGC~7070A and ESO
2400100 with the EPIC instruments (see Table~\ref{table2}).
We applied the standard reduction to the data mainly with SAS version
7.0 ({\tt http://xmm.vilspa.esa.es}).  We have further cleaned
the data set for periods of high background, as suggested by the
XMM-Newton Science Analysis system:Users' Guide available on 
line\footnote{\tt http://xmm.vilspa.esa.es/external/xmm\_user\_support/
documentation/sas\_usg/USG}.  We retained only patterns 0-4 inclusive for
EPIC-pn, and 0-12 for EPIC-MOS.  For imaging analysis, we have merged the
two MOS event lists to improve the statistics.  The merged event list will
not be used for spectral analysis.  

For an accurate determination of the background, and of the total
extent of the source, we have followed two parallel procedures:
1) we have produced radial profiles in different energy bands (see
later) and examined them to determine whether and when the profile
would become constant with radius and b) we compared them with the
equivalent photon distributions obtained from the blank sky fields
provided at {\tt http://xmm.vilspa.esa.es/external/xmm\_sw\_cal/
background/blank\_sky.shtml}.  These latter have been examined and cleaned
of residual high background levels to match the average value of our
cleaned fields.  Since the size of the sources is small compared to the
field of view, a local determination of the background is possible and
extrapolation to small radii is not limited by vignetting effects. We
therefore resolved to determine the background both for spatial and
spectral analysis from regions free of sources as close to the source
as possible.

To increase the statistics without having to take into account the different
patterns in the CCD-gaps in the two instruments, we have summed all EPIC-MOS
data and kept the EPIC-pn data separate. 
The limited statistics that result from these datasets do not allow us to
investigate in detail the X-ray characteristics of the two galaxies.  
Table~\ref{table3} summarizes the relevant X-ray measures for both galaxies.
 To convert the count rates into  fluxes and
luminosities we have assumed a conversion factor based on the EPIC-pn camera,
since  neither source is affected by the CCD gaps, and the XSPEC results
(see section 4.2).  

\subsection{Far UV observations}

Simultaneous  Optical Monitor  \citep[OM hereafter,][]{Mason01} 
images were obtained during the X-ray observations 
performed with the EPIC cameras
(see Table~\ref{table2}). The ultraviolet  UVW1 and UVM2 filters, 
which cover the ranges 245-320 nm and 205-245 nm respectively
\citep{Mason01},  and the U (300-390 nm)
and B (390-490) bands have been used. 
The Point Spread Function -- FWHM -- is 
$\approx$2.3\arcsec\ in U, $\approx$2.0\arcsec\ in B,
$\approx$1.7\arcsec\ UVW1, and 2.0\arcsec\ in UVM2 sampled
with 0\farcs 476$\times$0\farcs 476 pixels. 
Unwanted ghosts due to scattered light are unfortunately present in the OM images, 
less prominent when using UV filters with respect to
optical filters \citep[see][]{Mason01}. Accounting for the effects of
these features on the surface photometry is not an easy task. We simply mask
ghosts when they are away from our targets and do not affect 
the accuracy of the  photometry.  In more
difficult cases, when they could contaminate our estimates, 
we attempt to model such features in order to remove 
them from the image. 

In order to complete the UV information for our galaxies we
have also retrieved the {\it GALEX} data for NGC 474 (the only one
observed so far) from the archive.
The {\it GALEX} mission and instruments are fully described  in
\citet{Martin05} and \citet{Morrissey05,Morrissey07}. The spatial resolution 
of the images is $\approx$4\farcs 5 and 6\farcs 0 FWHM in FUV 
(135 -- 175 nm) and NUV (175 -- 275 nm) respectively, 
sampled with 1\farcs 5$\times$1\farcs 5 pixels.

The analysis on the surface photometry was carried out on the background
subtracted images we obtained from the archive, using the {\tt ELLIPSE}
fitting routine in the {\tt STSDAS} package of {\tt IRAF} and the {\tt
GALFIT} package \citep{Peng02}. The necessary AB photometric zero points
were taken from \citet{Morrissey05} for {\it GALEX} and from the {\it
XMM Users' Handbook}\footnote{\tt
http://xmm.vilspa.esa.es/external/xmm\_user\_support/
documentation/uhb/node75.html} for OM observations.  AB magnitudes,
photometric errors and colors were determined from the original data. 
Smoothed images are used to better enhance the faint shell structure,
mostly for displaying and for assessing the morphology.

{\tt ELLIPSE} computes a Fourier expansion for each successive isophote
\citep{Jedrzejewski87}, resulting in the surface brightness profiles in
the AB photometric system. {\tt GALFIT} was used to perform a bulge-disk
decomposition -- if needed -- and to determine the parameters of the
Sersic profile \citep{Sersic68} fitted  to the bulge of the galaxies. 
The profile is  sensitive to structural
differences between different kinds of early-type galaxies and thus
provides a better fit to real galaxy profiles.

\section{Results}

 \subsection{X-ray morphology}

To study the distribution of the X-ray emission associated with these
galaxies we have produced smoothed images, for a visual impression of
the X-ray morphology,  and radial profiles, that measure the extent and
possible azimuthal asymmetries.

The smoothed X-ray maps of the emission in two broad energy bands are
shown in Fig.~\ref{fig1}, superposed onto the optical images from the DSS
plate.  We used the $csmooth$ adaptive smoothing algorithm that takes
into account the available statistics to determine the size of the
kernel to be applied \citep[see][]{Ebeling06}. Both galaxies are
clearly detected, and the emission is centered on the nuclear region. Several unrelated sources are also present in the field, in both observations.  Note in particular a relatively bright compact source to the SE of NGC~7070A.
It is immediately evident that 
the emission from these two galaxies has significantly different
characteristics, even though caution should be used when interpreting
adaptively smoothed images.  
In NGC~7070A both hard and soft photons are present,
but the emission appears to be rather compact
(a more quantitative measure is discussed later).  In ESO~2400100 the
soft band emission is significantly stronger than the hard band one,
and appears broader than in NGC~7070A.

The radial profiles of the emission shown in  Fig.~\ref{fig2} and
Fig.~\ref{fig3} confirm
the visual impression given by the maps (Fig.~\ref{fig1}).  
Figure~\ref{fig2} shows NGC~7070A  in the soft and hard broad energy bands. 
In Fig.~\ref{fig3} for ESO~2400100 only in the soft band is shown. 
All unrelated/background sources have been excluded from the profiles. 
The photon distributions in the 0.3-3.0 keV soft band are centered
onto the X-ray peak and fitted with 
a $\beta$-model of the kind: $\Sigma_x \approx (1+({r \over r_c})^2)^{-3\beta
+ 0.5}$ in both galaxies.  However, in NGC~7070A, 
the best fit model  is entirely consistent with the instrument 
Point Spread Function (PSF), which can be reasonably parameterized with
a $\beta$-model with $\rm r_c=5''-6''$  (EPIC-MOS and $pn$ respectively) and
$\beta=0.63$ (see the
XMM-SOC-CAL-TN-0029 and XMM-SOC-CAL-TN-0022 documents online at {\tt
http://xmm.vilspa.esa.es}). The fit quality is better in the harder 
3.0-7.0 keV energy band, most likely a consequence of the presence of
absorption that reduces transmission at softer energies (see later). 
In ESO~2400100, the $\beta$-model indicates a core radius of $8''$,  larger
than the PSF.  We have then used a $\beta$-model convolved with the 
EPIC PSF, 
and we find a core radius of $\rm  4''$, corresponding to
$<1.0$ kpc.  No radial profile is given for
the hard energy band in ESO~2400100 for lack of emission above $\sim 3-4$ keV.

\subsection{X-ray  spectral properties}

In order to obtain the spectral parameters for the two galaxies, we
have used regions selected to maximize the signal and avoid the CCD
gaps (in the EPIC-pn data).  The data have been rebinned in larger
energy channels to increase the statistics, and have been background
subtracted, with the background data chosen locally in nearby regions
where the galaxy's emission is no longer detectable as suggested by
the radial profiles.  We have used  a thin  plasma and a power law models,
with low energy absorption to parametrize the spectral shape, assuming
the Wilms et al. (2000) abundance tables for the gas metallicities.  In all
cases the low energy absorption is at least the value corresponding to
the line-of-sight Galactic N$_H$ \citep[3.1 and 1.7 $\times$ 10$^{20}$
cm$^{-2}$, for NGC 7070A and \eso\ respectively,][]{dickey}. 
Luminosities are corrected for absorption and are quoted in the 0.5-2.0 keV band for the soft plasma component, and in the broad 0.5-10 keV band for the power law component (representing the contribution from
either the AGN or the binary sources). 

\begin{figure*}
\resizebox{18cm}{!}{
\includegraphics[width=8.3cm]{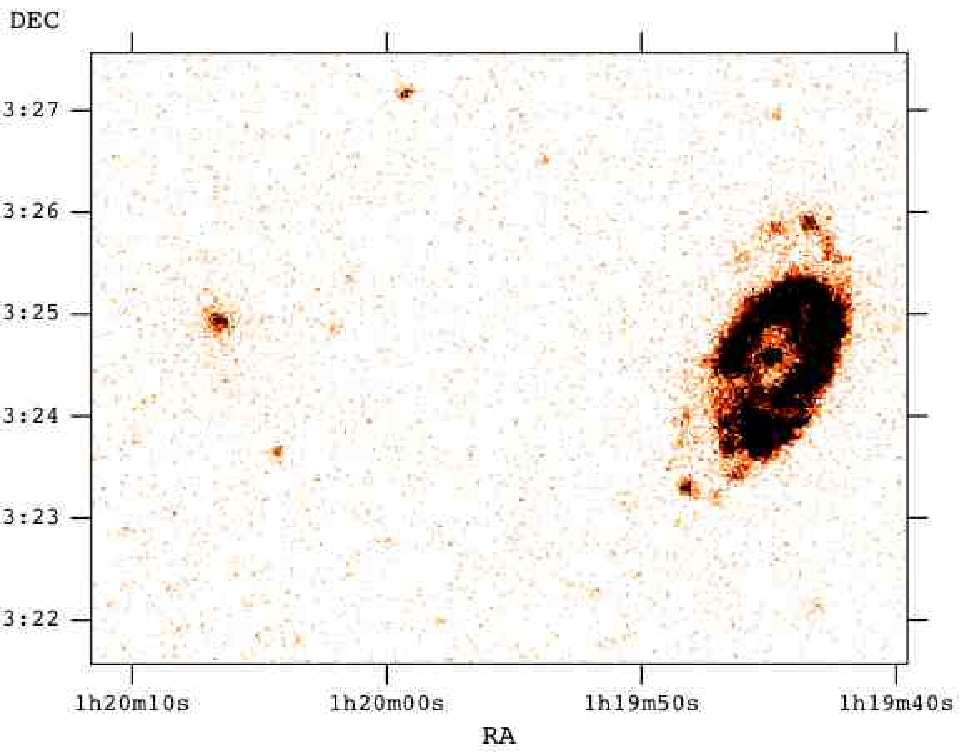}
\includegraphics[width=8.3cm]{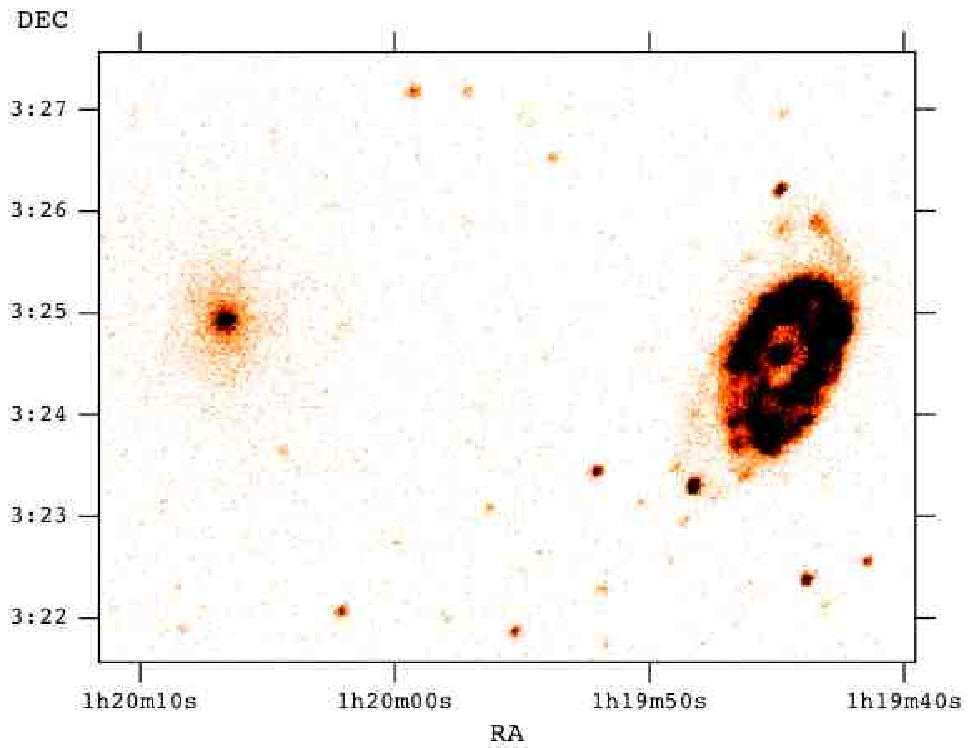}}
{\includegraphics[width=8.8cm]{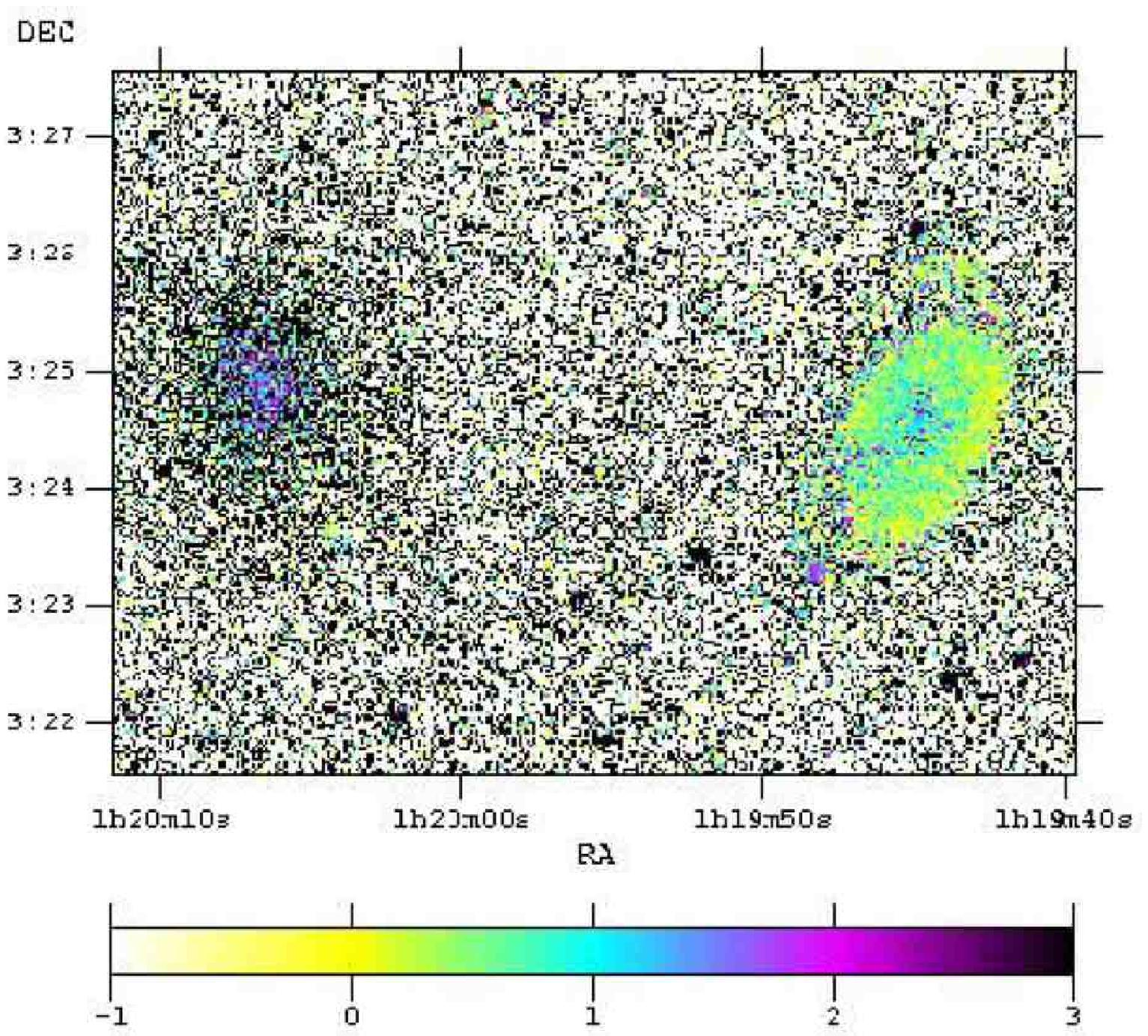}
\includegraphics[width=8.0cm]{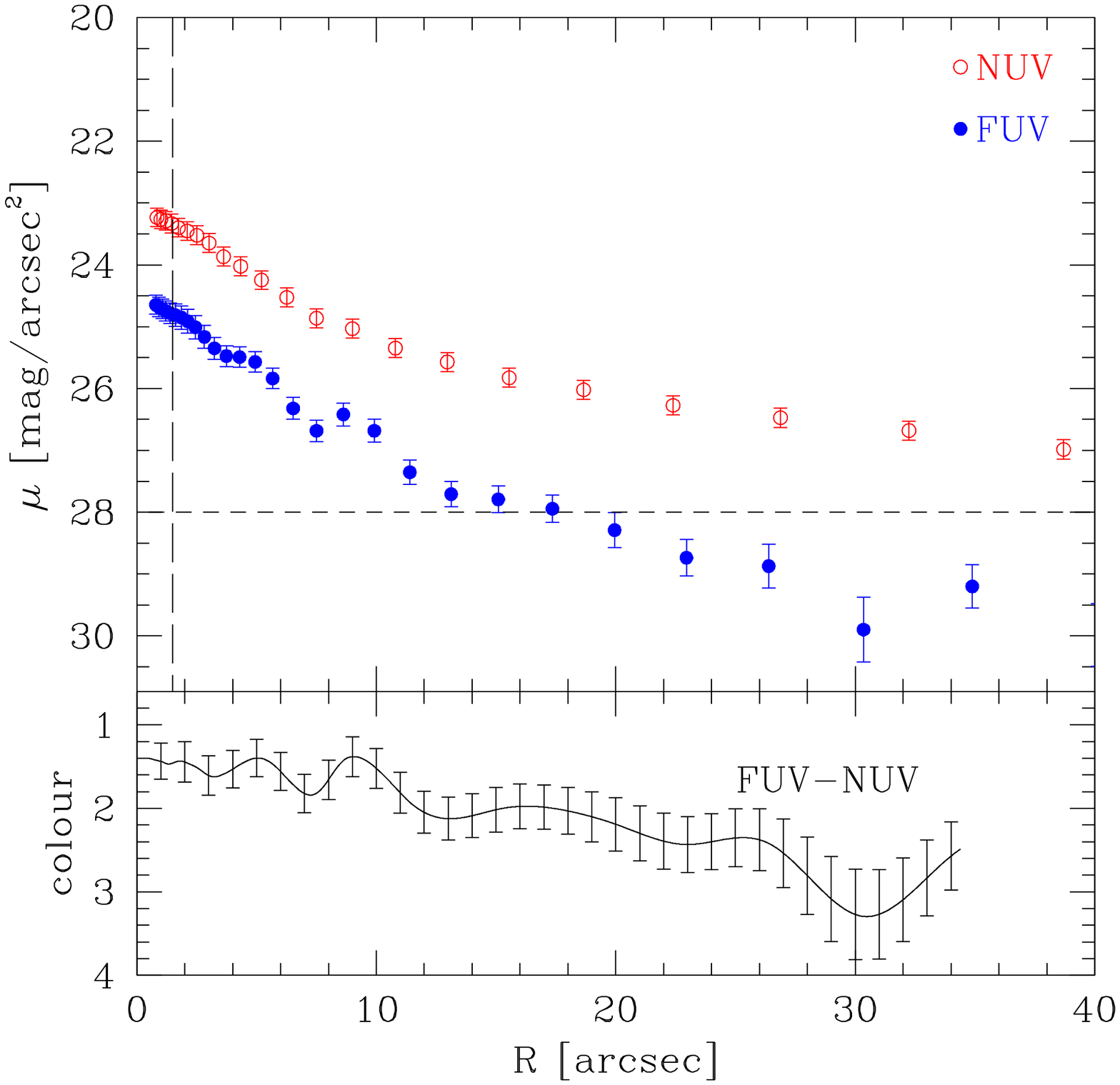}}
\caption{NGC 470/474 full resolution {\it GALEX} FUV ({\it top left}) and NUV
({\it top right}) background subtracted images  in counts~px$^{-1}$~s$^{-1}$.
Pixel by pixel {\it GALEX} (FUV-NUV) 2D color map of the NGC 470/474
system ({\it bottom left}). FUV and NUV luminosity and (FUV-NUV) color profiles of 
NGC~474 ({\it bottom right}). }
\label{fig5}
\end{figure*}

\underbar{NGC~7070A}~~ The full spectral range, from $\sim 0.3$ to
$\sim 10$ keV, is covered with adequate signal for analysis. The spectrum
is described by a combination of a highly absorbed and a significantly
fainter unabsorbed power law, with $\Gamma \sim 1.5$ and N$\rm _H \sim
2 \times 10^{23}$ cm$^{-2}$, corresponding to the typical
parametrization of an absorbed AGN (see Fig.~\ref{fig4}).  The intrinsic
luminosity of the source, corrected for absorption, is L$_X$=7 $\times$
10$^{41}$ \ergs\ (0.5-10.0 keV), indicative of a moderate AGN. 
Notice that NGC~7070A was undetected both in \hi\ and in the
radio continuum; \citet{Oosterloo02} provide an upper limit of 
M$_{HI}< 0.5 \times 10^8$ M$_{\odot}$ giving  M$_{HI}/L_B <$0.003.
While the data do not require a plasma component, we can place an upper
limit to a contribution from a hot ISM to  L$_X<$2 $\times$
10$^{38}$ \ergs\ (0.5-2.0 keV), for a temperature of 0.3 keV and  metal
abundance at 30-100\% solar.

\underbar{ESO~2400100}. The emission can be parameterized by a low temperature 
plasma (kT $\approx$ 0.3-0.5) with a minor contribution from a power law
component with $\Gamma \sim 1.7$ to account for residuals at 
high energies (Fig.~\ref{fig4}). A small amount of absorption of 
N$\rm _H \sim$ 4 $\times$ 10$^{20}$ cm$^{-2}$, in addition  to the line 
of sight Galactic value, is also suggested by the data, 
to contain the power law contribution at low energies.  
The intrinsic luminosity for the gas component
is $\rm L_X \sim5 \times $ 10$^{39}$ \ergs\ (0.5-2.0 keV),
while the integrated emission from the power law component is 
$\rm L_X \sim 1\times $  10$^{40}$ \ergs\ (0.5-10 keV).

\begin{figure*}
\resizebox{18cm}{!}{
\includegraphics[angle=-90]{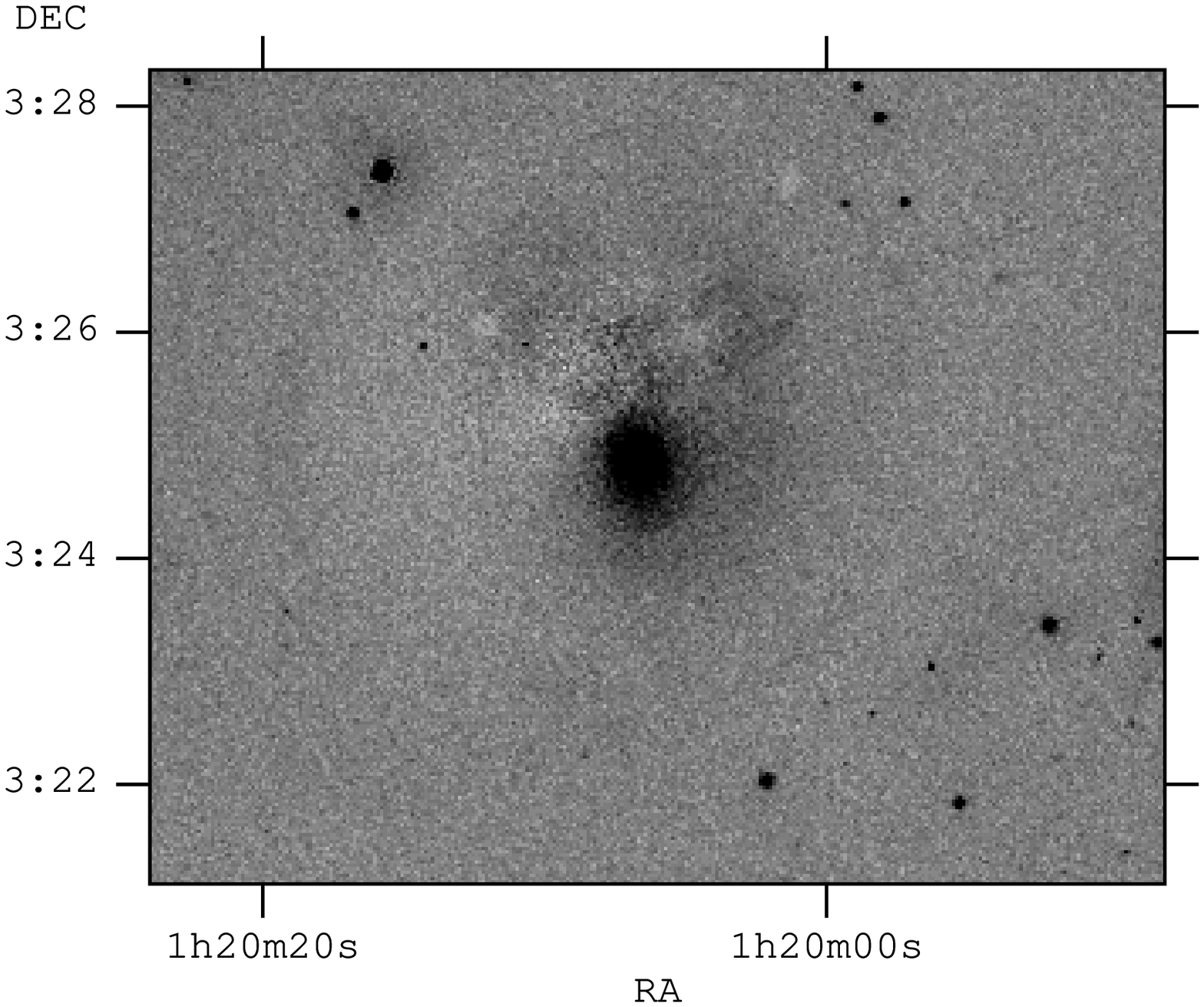}
\includegraphics[angle=-90]{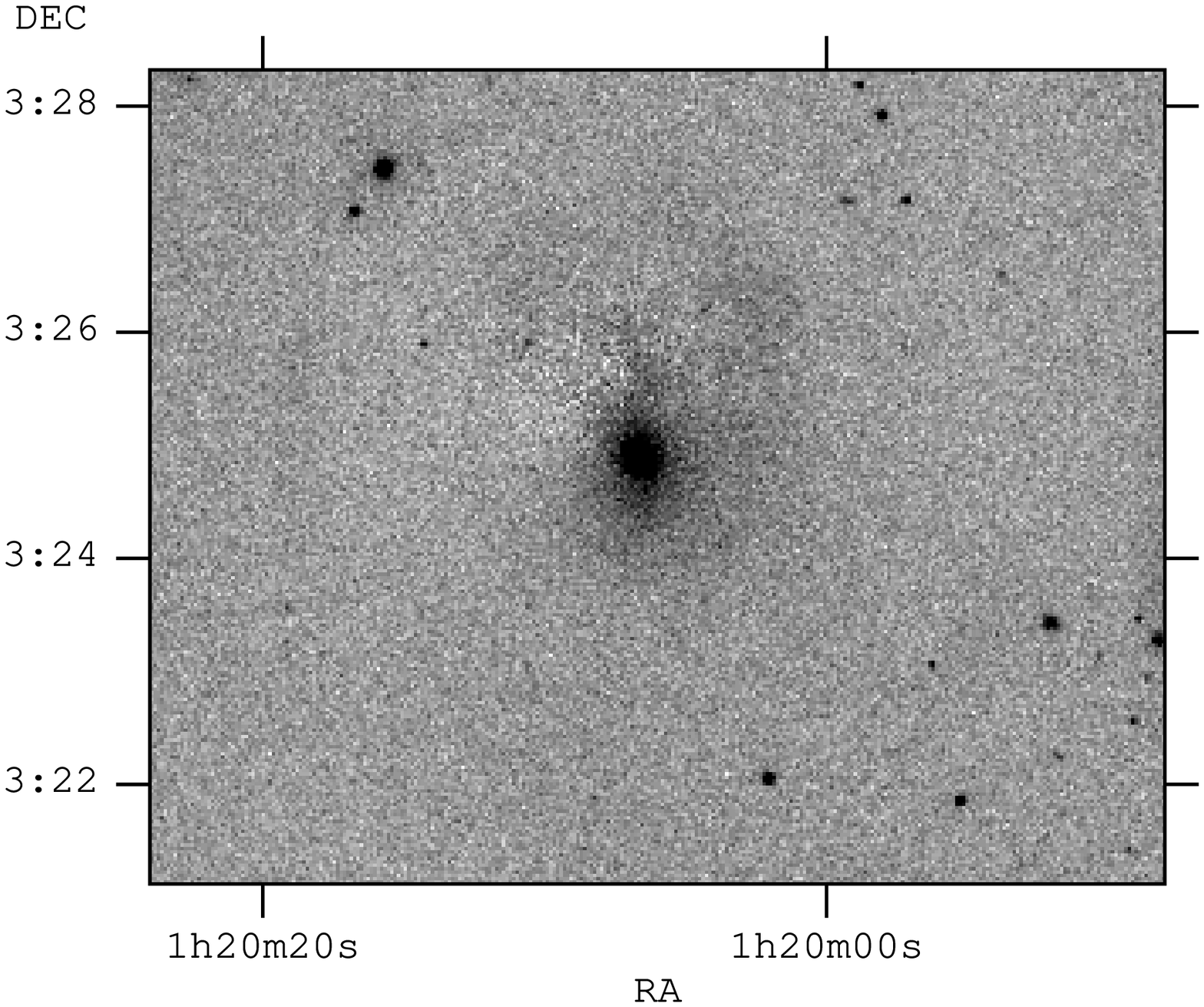}
\includegraphics[angle=-90]{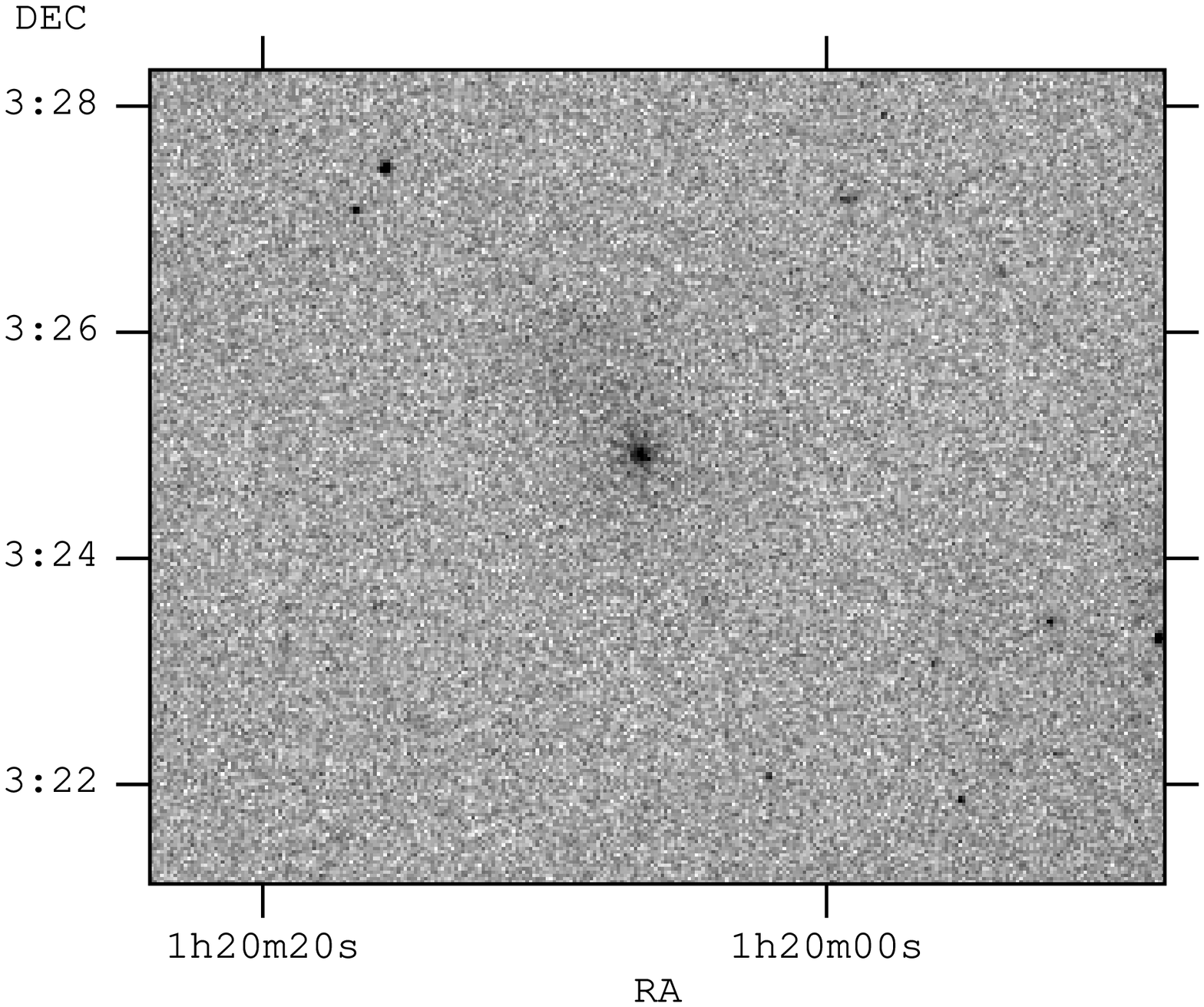}
}
\caption{NGC 474  XMM-OM in U ({\it left }), UVW1
({\it mid }) and UVM2 ({\it right }). The presence of  
ghosts, north of the nucleus, which we attempt to correct, 
perturbed the detection of inner shell in the U-band and 
UVW1 images. The shells south of the nucleus and the 
outer (north east and south west) shells are still 
visible in the U-band and UVW1 images.
See for comparison \citet{Pierfederici04} Figure~1. }
\label{fig6}
\end{figure*}
\begin{table*}
\caption{Summary of the X-ray and the UV photometry}
\label{table3}
\tiny
\begin{tabular}{lcccc}
\hline
\hline
{\bf X-ray data}                                                     &             & &          \\
 & NGC~474 &  NGC 7070A& ESO 2400100 \\
\hline
PN counts         &   75.2$\pm$15  &2530$\pm$55 & 650$\pm$40   \\
count rate [cts s$^{-1}$]     &   1.2 $\times$ 10$^{-2}$
&9.7 $\times$ 10$^{-2}$ & 3.3 $\times$ 10$^{-2}$   \\
log L$_X$-gas (0.5-2 keV)  [erg s$^{-1}$] &-- & $<38.30$&39.70 \\
log L$_X$-pow (0.5-10 keV)  [erg s$^{-1}$] & -- 
                                           &41.85 & 40.00 \\
log L$_X$-tot (0.3-6 keV)  [erg s$^{-1}$] & 39.60 & [41.70] & 40.11     \\
log L$_B$  [erg s$^{-1}$ L$^{-1}_{\sun}$] & 10.27 & 9.86  & 10.39   \\
log L$_K$  [erg s$^{-1}$ L$^{-1}_{\sun}$] & 10.97 & 10.72 & 11.14   \\
log L$_X$/L$_B$ [erg s$^{-1}$ L$^{-1}_{B\sun}$] (0.3-6 keV) & 29.35 & [31.84] & 29.72\\
\hline
{\bf Far UV and optical data}                                              &          & &            \\
                                        & NGC~474             & NGC 7070A & ESO 2400100 \\
\hline             
r$_{e}^{NUV}$ [arcsec]       &         17                 &          & \\
r$_{e}^{FUV}$ [arcsec]       &          15               &          & \\
r$_{e}^{B}$ [arcsec]           &                             &74.68$\pm$0.67 & \\
r$_{e}^{U}$ [arcsec]           &      38.11$\pm$0.85  &          & 118.13$\pm$2.37\\
r$_{e}^{UVW1}$ [arcsec]     &     45.64$\pm$1.34   &          & 54.53$\pm$1.12 \\
r$_{e}^{UVM2}$ [arcsec]     &      50.03$\pm$4.94   & 21.63$\pm$4.81& 9.20$\pm$0.44  \\
$n_{Sersic}$ [U]               &       3.66$\pm$0.04    & 2.10$\pm$0.01 &  4.47$\pm$0.04 \\
$n_{Sersic}$ [UVW1]          &      3.75$\pm$0.06    &          & 3.39$\pm$0.03\\
$n_{Sersic}$ [UVM2]          &      3.18$\pm$0.14    & 0.69$\pm$0.12  & 1.11$\pm$0.05  \\
m$^{tot}_{FUV}$               & 19.08 $\pm$ 0.05  & & \\    
m$^{tot}_{NUV}$               & 16.53 $\pm$ 0.01 &  & \\
m$^{tot}_{V}$                   &  11.46                    &  12.29$\pm$0.14  &  11.75\\
$\langle$(UVW1-U)$\rangle$ &  1.01$\pm$0.09     &                      &   1.13$\pm$0.10\\
$\langle$(UVM2-UVW1)$\rangle$ &2.07$\pm$0.13 &                      &   1.83$\pm$0.31\\
$\langle$(UVM2-U)$\rangle$ & 3.07$\pm$0.18      &                      &   3.00$\pm$0.30 \\
$\langle$(FUV-NUV)$\rangle$ &   2.56$\pm$0.05   &                      & \\
$\langle$(UVM2-B)$\rangle$ &                            & 4.63$\pm$0.06 & \\
\hline
\end{tabular}

\medskip
Notes:  
A full description of the X-ray data of NGC~474 is provided in \citet{Rampazzo06}. X-ray luminosities for different components, described in
section 4.2, are given in the appropriate X-ray bands.  The  total emission is given in 0.3-6.0 keV band for comparison with \citet{Brassington07}, and will be used in Fig.~\ref{fig11}. Note that the value quoted for NGC 7070A refers to the emission of the nuclear source, as measured in section 4.2. In Fig.~\ref{fig11} we use an estimate of the contribution from the galaxy, as we explain in Section 5.2, which gives log L$_X$/L$_B$=29.84.
NUV and FUV bands refers to the {\it GALEX} observations as described
in Section~2; B, U, UVW1, UVM2 bands refers to XMM-OM observations.
The photometric data of ESO~2400100 refers to the entire galaxy envelope, 
i.e. includes the two detected nuclei. 
AB magnitudes have been corrected for Galactic extinction.
We adopt the following formulae for our galaxies \citep[see ][]{Brassington07}:
log $L_B$ ($L_\odot$) =12.192 -0.4$B_T$ +2log$D$ \citep{Tully88} 
where $B_T$ and $D$ are the apparent magnitude and the
distance in Mpc; $L_K$ = 11.364 -0.4$K_T$ + log(1+$z$) - 2log$D$ 
\citep{Seigar05}, $K_T$ is the K-band magnitudes taken from the 2MASS survey, $z$
is the galaxy redshift, and $D$ the distance in Mpc.
\end{table*}

   \begin{figure*}
      \resizebox{18cm}{!}{
   \includegraphics[angle=-90]{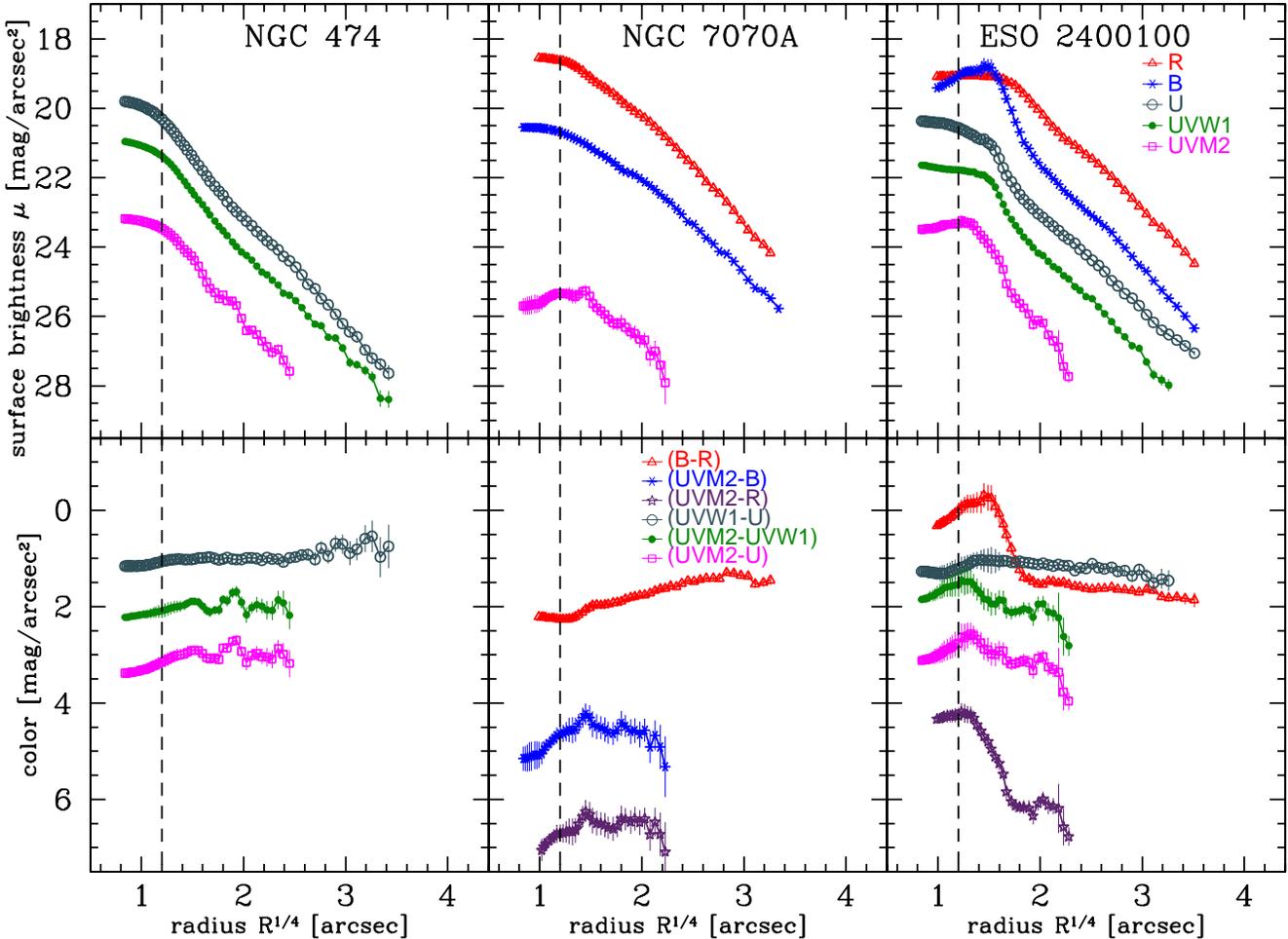}}
    \caption{Luminosity and color profiles ({\it top and bottom panels
    respectively}).  The vertical dashed line represent the OM Points
    Spread Function (FWHM) in the UVW1 band (see text). The R-band 
    photometry of NGC~7070A and ESO~2400100 is based on
    \citet{Lauberts89} calibrated data  with a spatial resolution
    sampled by 2\arcsec $\times$ 2\arcsec pixels. The B, U, UVW1 and
    UVM2 luminosity profiles are obtained from the present OM
    observations. The two nuclei embedded in the body of ESO~2400100 are
    separated by $\approx$5\arcsec. The center of the surface photometry
    is set by the outer isophotes and it is located 2\farcs 85 north
    west of the southern nucleus (we indicated as ESO~2400100a) along
    the line connecting the   two nuclei. The northern nucleus,
    ESO~2400100b, is visibly bluer than the southern one.}
         \label{fig9}
   \end{figure*}

The spectral results are in substantial agreement with the picture
suggested by the spatial analysis presented above:  the nuclear source is
the dominant component in NGC 7070A. 
In ESO~2400100, the emission is indicative of a low temperature plasma
with a contribution from the collective emission of individual sources
in the galaxy.  Given the lack of evidence of a central
source from the radial profile, and of an AGN in the optical data,
the alternative of a mildly absorbed,
low luminosity central source is less likely, although not entirely excluded.

\subsection{ Far UV morphology and photometry}

We performed a morphological study of the galaxy structures
both with the OM and {\it GALEX} data. 

The photometric and structural values derived from the present 
surface photometry are summarized in Table~\ref{table3}. 
Notice that the large variation both in the effective radii and Sersic 
indices  measured in the UV and optical bands studied are 
partly indicative of the different S/N of images in the different bands. 

Figure~\ref{fig5} shows the {\it GALEX} FUV (top left panel) and NUV
(top right panel) images, obtained from the archive. 
The morphology of the spiral galaxy NGC 470 is similar in 
the FUV and NUV bands, while only the very inner part of NGC 474 is visible
in either of them, consistent with the
difference in their respective morphologies.
In Figure \ref{fig5} (bottom left) the 
{\it GALEX} (FUV - NUV)  colour map also illustrates the very different results
obtained for the two galaxies in the pair, and 
shows a very young stellar component only in NGC~470.
In the bottom right panel we show the NUV and FUV 
luminosity profiles as well as the color profile of NGC~474. The color profile 
shows a gradient going from the center, ($\langle FUV-NUV \rangle \approx$1.5) to
the outskirts of the galaxy ($\langle FUV-NUV \rangle \approx$2.5).

Although the effect could be partly due to the lower
S/N of the FUV with respect to the NUV image, 
the FUV emission seems more concentrated
in the central part of NGC~474 with respect to the NUV emission. 
There is no evidence of the complex shell structure detected 
by \citet{Pierfederici04} and \citet{Sikkema07}, which is instead
present in the  U and UVW1 OM images shown in Figure~\ref{fig6}.
Ghosts perturb the shape of the shells to the north of the nucleus but there is a one to one 
correspondence to the shell system evidenced in Fig.~1 of
\citet{Pierfederici04}. Asymmetries in the galaxy nucleus could be 
attributed to the presence of dust shown also by the HST-ACS
image in \citet{Sikkema07}.

In Figure~\ref{fig9} (left panels) we plot the luminosity 
and the color profiles of NGC~474. The galaxy is considered a {\it bona fide} S0
seen nearly face-on: we measured an average ellipticity of $\approx$0.3. 
The Sersic index $n$ is quite stable around the value of 3.5 in the U, UVW1 and 
UVM2 (see Table~\ref{table3}) while the effective radius r$_e$ shows a 
significant variation which is partly due to the presence of ghosts north
of the nucleus.  We do not detect radial color gradients within the errors 
in all bands analyzed,  at odds with the slightly decreasing {\it GALEX} (FUV-NUV) color profile.
Our average (UVW1-U), (UVM2-UVW1) and (UVM2-U) color measures 
listed in Table~3 agree within the errors with the global color of the galaxy 
provided by the OM source-list catalogue which reports a
 total galaxy magnitude of 16.14 and  17.21
and 19.30 in the U, UVW1 and UVM2 bands respectively.

\begin{figure}
\resizebox{9cm}{!}{
\includegraphics[angle=-90]{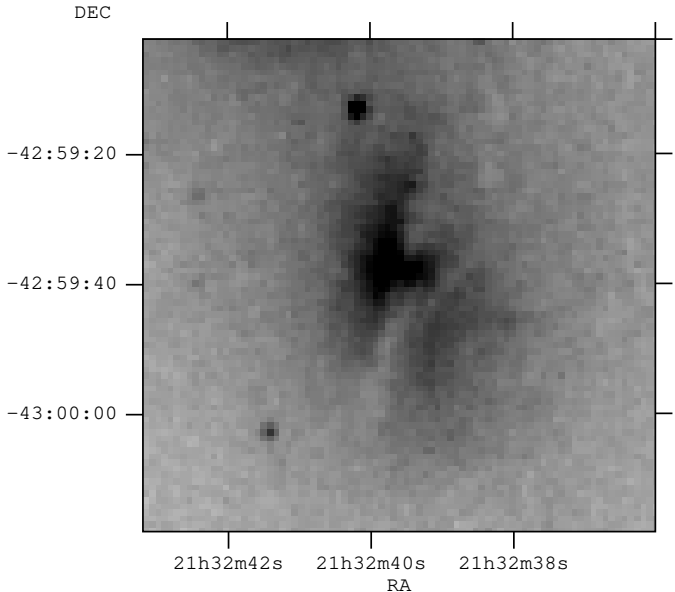}
}
\caption{OM B-band  image of  NGC~7070A.
The complex system of dust crossing the entire galaxy 
is clearly visible in the B-band. 
}
 \label{fig7}
   \end{figure}

Figure~\ref{fig7} shows the B image of NGC~7070A. The dust lane is
clearly visible. Although the exposure time is similar 
to that of the other galaxies in the sample, the galaxy is
barely visible in the UVM2 band (hence we do not show it), an effect probably due to dust
absorption, more significant than in the other objects.  The luminosity 
profiles in the B and  UVM2 bands are shown in Figure~\ref{fig9}
together with the R-band luminosity profile obtained from the ESO-Uppsala
data base. The shape of the luminosity profiles indicates the presence of 
two components, probably a bulge plus a disk, as suggested
by the kinematic analysis of \citet{Sharples83}.
 The (UVM2-B) and (UVM2-R) color profiles are flat.
The (UVM2-B) color reported in Table~3 as the average over the profile in Fig~\ref{fig9}
agrees within the
errors with the global color of the galaxy provided by the OM source-list
catalogue, which also gives a total galaxy magnitude of 16.20 and  20.66 in the
B and UVM2 bands respectively. 

\begin{figure*}
   \resizebox{18cm}{!}{
   \includegraphics[angle=-90]{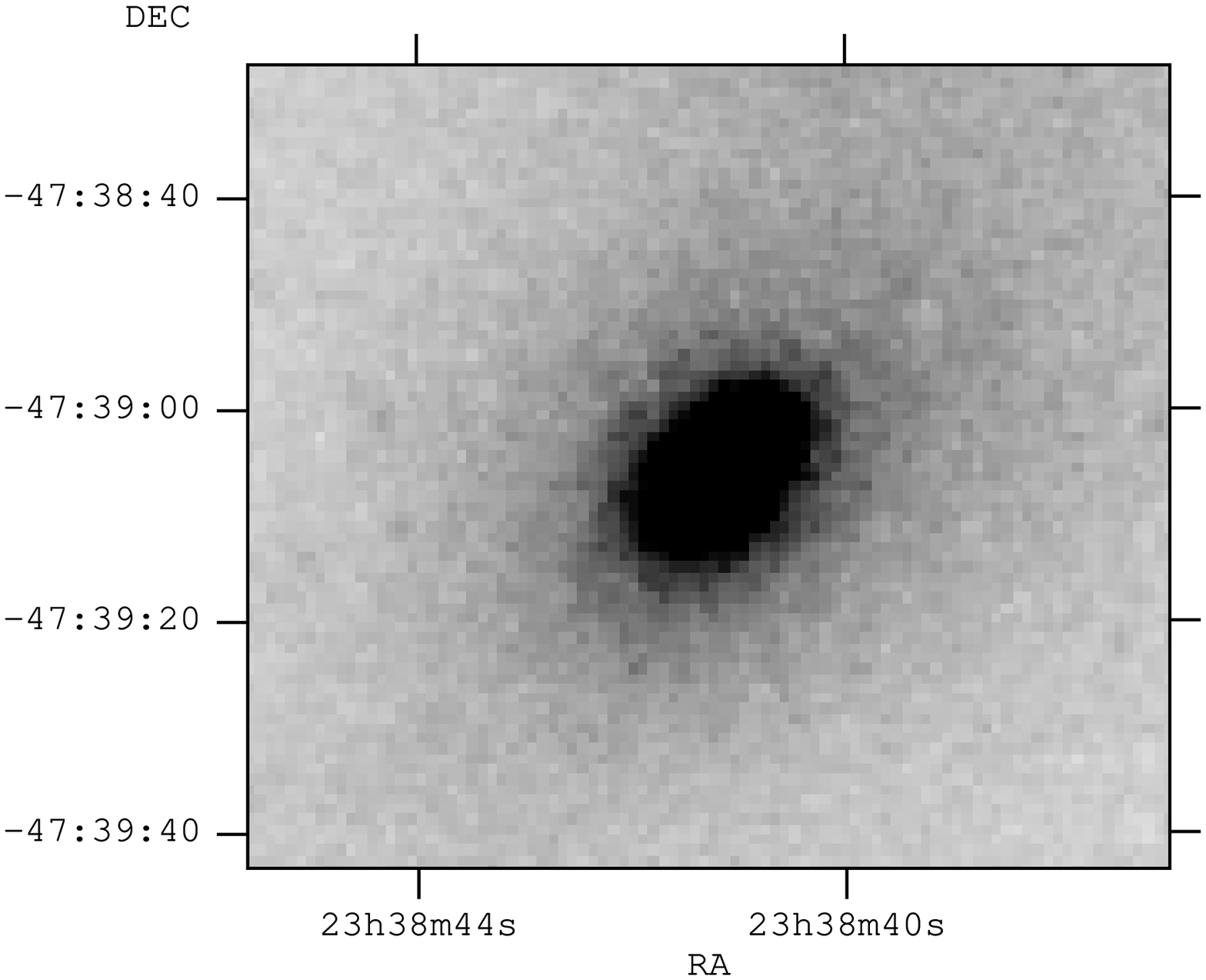}
   \includegraphics[angle=-90]{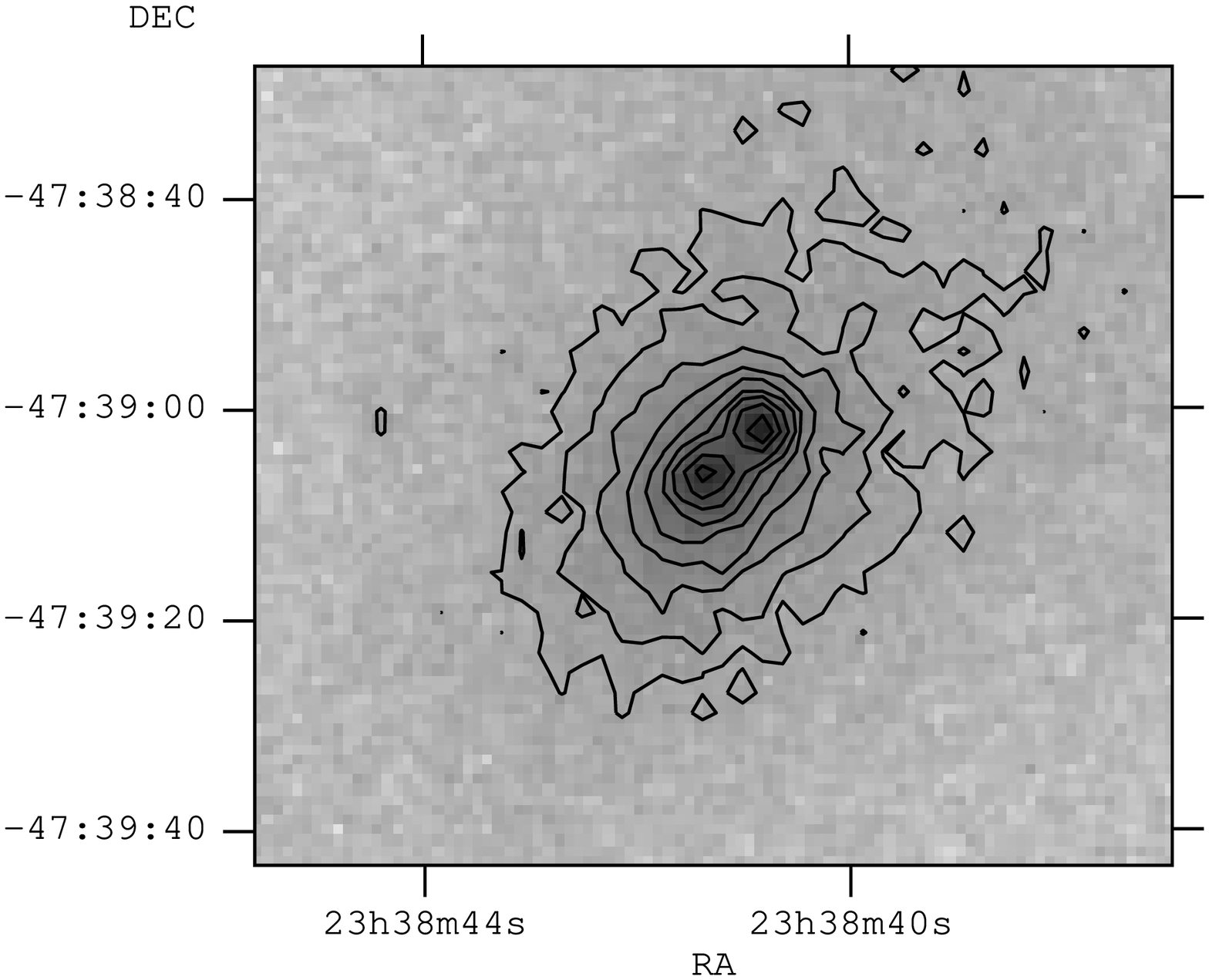}
   \includegraphics[angle=-90]{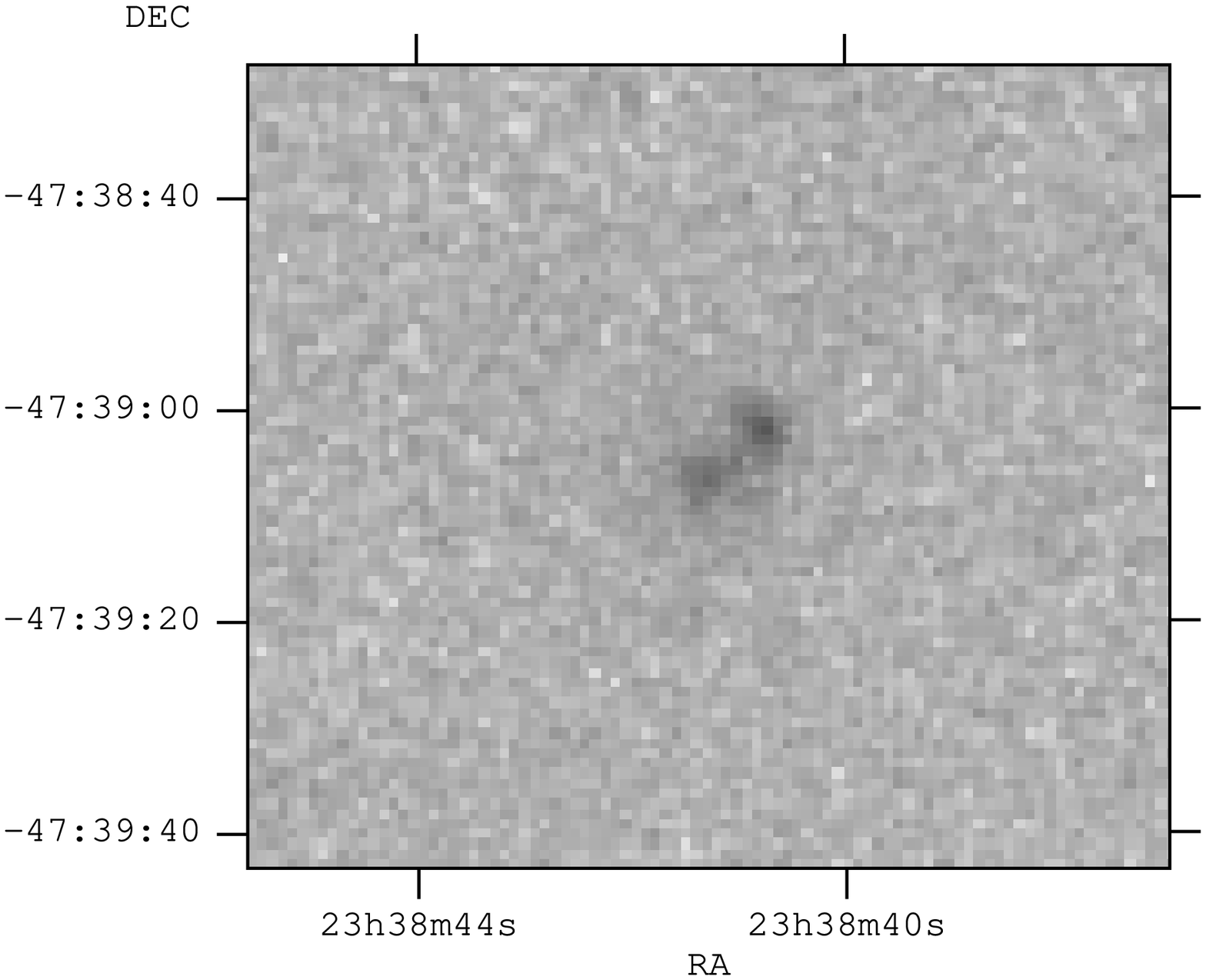}}
    \caption{U ({\it left panel}), UVW1 ({\it mid panel}) and UVM2 ({\it right panel}) 
images of ESO~2400100. In the mid panel we plot the isophotes to show the structure 
of the galaxy. The two nuclei (we named ESO~2400100$a$ the southern
nucleus, ES0~2400100$b$ the northern one) embedded in the galaxy body are 
clearly visible in the UVW1 and UVM2 images.}
         \label{fig8}
   \end{figure*}

Figure~\ref{fig8} shows the U, UVW1 and UVM2 images of ESO~2400100.
The two nuclei, discovered by the kinematical study of \citet{Longhetti98a},
 are clearly visible also in the UVW1 and UVM2 bands. Based on 
kinematical measures and models by \citet{Combes95}, \citet{Longhetti98a} 
suggest that they are strongly interacting and not a mere projection effect. 

In  Fig.~\ref{fig9} we plot the surface photometry in several bands and
the color gradients.  B and R-bands luminosity profiles obtained from the
ESO-Uppsala data base are also shown.  The average colors of the galaxy
are given Table~3.    The present surface photometry considers the galaxy
as a single envelope.  The geometric center of the galaxy, as determined
by the outer isophotes, is located between the two distinct nuclei.
This explains why the shape of the light distribution in the B-band
profiles shows a significant rise in the center, at about 2\farcs 85, or
a  substantially flat distributions in the other bands, and a flattening
outside when the two distinct components disappear and the galaxy
becomes a unique envelope.  As already noticed in NGC 7070A above, the
luminosity profile also suggests the need for a disk component.  Except
for (UVW1-U), all color profiles in Figure~\ref{fig9} show a gradient
from the center to the outskirts, more evident in  (B-R) and (UVM2-R).
The OM source-list catalogue provides  the following magnitudes for the
Northern nucleus: U=16.31, UVW1=17.38, UVM2=19.03. For the Southern one:
U=16.35, UVW1=17.37, UVM2=19.15. Therefore we can estimate different UV
colors for the two nuclei:  (UVW1-U)$_{North}$ = 1.07 and
 (UVW1-U)$_{South}$ = 1.02;  (UVM2-U)$_{North}$ = 2.71 and
  (UVM2-U)$_{South}$ = 2.79; (UVM2-UVW1)$_{North}$ = 1.64 and
  (UVM2-UVW1)$_{South}$ = 1.77; these colors are quite similar, also
considering the errors reported in Table~3. A very blue value of (B-R)
= -0.31 is derived for the Northern nucleus from the ESO-Uppsala data.

\section{Discussion}

We use the results obtained above to investigate whether the 
combination of the far UV and optical colors can provide useful
constrains on the time at which the accretion/merging phenomenon has occurred.
Possible 
{\it rejuvenation} signatures in the stellar population generated during the shell 
formation have already been traced and the galaxy secular evolution has been dated 
through {\it GALEX} observations  by, e.g., \citet{Rampazzo07}. 

This information is also relevant for a proper discussion
of the X-ray data. For example, the X-ray luminosity is thought to evolve
with different phases of the merging processes, from 
the galaxy-galaxy encounter, the merging itself and the coalescence 
of the remnant towards a 
relaxed galaxy \citep[see e.g.][]{Brassington07}.    

\subsection{Can we time the accretion event through modelling of {\it GALEX} 
and OM magnitudes and colours ?}

\citet{Longhetti00} studied the H$\beta$ vs. [MgFe] plane with the aim
of inferring the age of the stellar population in shell galaxies.
In the above plane, shell galaxies occupy the same region of {\it normal}
early-types, although a number of them have a  stronger H$\beta$. 
Given the similarity with shell galaxies, for which the signatures of
interaction are evident, \citet{Longhetti00} suggest that the scatter 
in the H$\beta$ vs. [MgFe] plane could be the result of a secondary
episode of star formation after their formation.
In fact the shape of the distribution of the
line-strength indices in the H$\beta$ vs. [MgFe] plane suggests an
effect related to the metal enrichment that always accompany star
formation. \citet{Longhetti99} further suggest that, if the last star
forming event is connected to the formation of the shell system as
expected from simulations, the shells ought to be a long lasting
phenomenon, since star forming events that occurred in the nuclear
region of shell galaxies are statistically old (from 0.1 up to several
Gyr). A subsequent study by \citet{Tantalo04b} 
added an additional parameter to interpret the scatter in the
H$\beta$ emission, i.e.  the 
enhancement in $\alpha$ elements in the abundance patterns. They
argue that part of the scatter in H$\beta$ could reflect the
intrinsic variation of $\alpha$ enhancement from galaxy to galaxy
existing in old populations of stars due to different star formation
histories rather than a dispersion in age caused by more recent star
forming episodes. 

Based on these premises, \citet{Rampazzo07} investigated whether {\it
GALEX} data may confirm or disprove the case that shell galaxies hosted
a recent star burst event. UV colours are considered to be particularly
suited to identify very young stellar populations and their distribution
inside interacting galaxies \citep[see e.g.][]{Hibbard05}. There is,
however, an important point to keep in mind dealing with UV colours,
i.e. the possibility that high metallicity, old stars (age $\geq
10^{10}$ yr and metallicity $Z\simeq 3Z_\odot$) likely present even in
small percentages in early-type galaxies could affect the (FUV-NUV)
colours. The evolved stars in metal rich populations (Extremely HB and
AGB-manqu\'e) are the proposed source of the UV excess in quiescent
early-type galaxies and any modeling of star formation must take this
population into consideration \citep[see][for a detailed discussion of
the subject]{Bressan94}. Fortunately, additional information,  such as
the presence of dust and gas, the dynamics or the overall spectral
energy distribution, can help  in a correct interpretation of  the
results.

The key assumptions in the definition of the stellar models and isochrones, the library 
of stellar spectra and the AB mag photometric system 
that are used to calculate theoretical
{\it GALEX} FUV and NUV magnitudes for a Simple Stellar Population
(SSP) are provided in \citet{Rampazzo07}.
We use them in modeling the {\it GALEX} data set for NGC~474.
The same assumptions have been adopted in modelling the B, U, UVW1 and UVM2
filters used for  NGC~7070A and ESO~2400100 with the
XMM-Newton Optical Monitor. 
Filter transmission curves, provided by the XMM - Newton Science Center (see 
also \S~3.2), have been elaborated to calculate magnitudes as in the  case of 
the {\it GALEX} filters.

The comparison of data with theory is made assuming that the complex
stellar mix of a real galaxy can be reduced to a SSP of suitable
metallicity and age. However this approximation has different
implications when interpreting the two parameters. While the metallicity
distribution can be ``reasonably'' approximated to the mean value,
the same does not hold for the age, when it is derived from
integrated properties \citep[see e.g.][]{Serra07}. In the discussion
below, one has to keep in mind that the age  measured
from colours (and/or indices) is always biased (dominated) by the last episode of
star formation. In other words, it is a mean luminosity weighted
age, in which the most recent star forming episode dominates at
least during the first 2-3 Gyr from its occurrence. This is simply
due to the rather well established law of luminosity fading of
stellar populations, which ultimately mirrors the lifetime and
evolutionary rate of a star as a function of its mass. In
the following we will refer to this age as $T_{SF}$. It does not
necessarily coincide with the real age of a galaxy $T_G$. In
general, if a galaxy underwent an initial star forming episode 
followed by quiescence, $T_{SF}\simeq T_G$; if episodes
of star formation occurred at a later time, $T_{SF} < T_G$, the difference
becoming
larger as the  activity is closer to us in time. An old
galaxy may look young if a secondary star formation activity took
place in the "recent past". 

   \begin{figure*}
         \resizebox{18.5cm}{!}{
     \includegraphics[width=7.5cm]{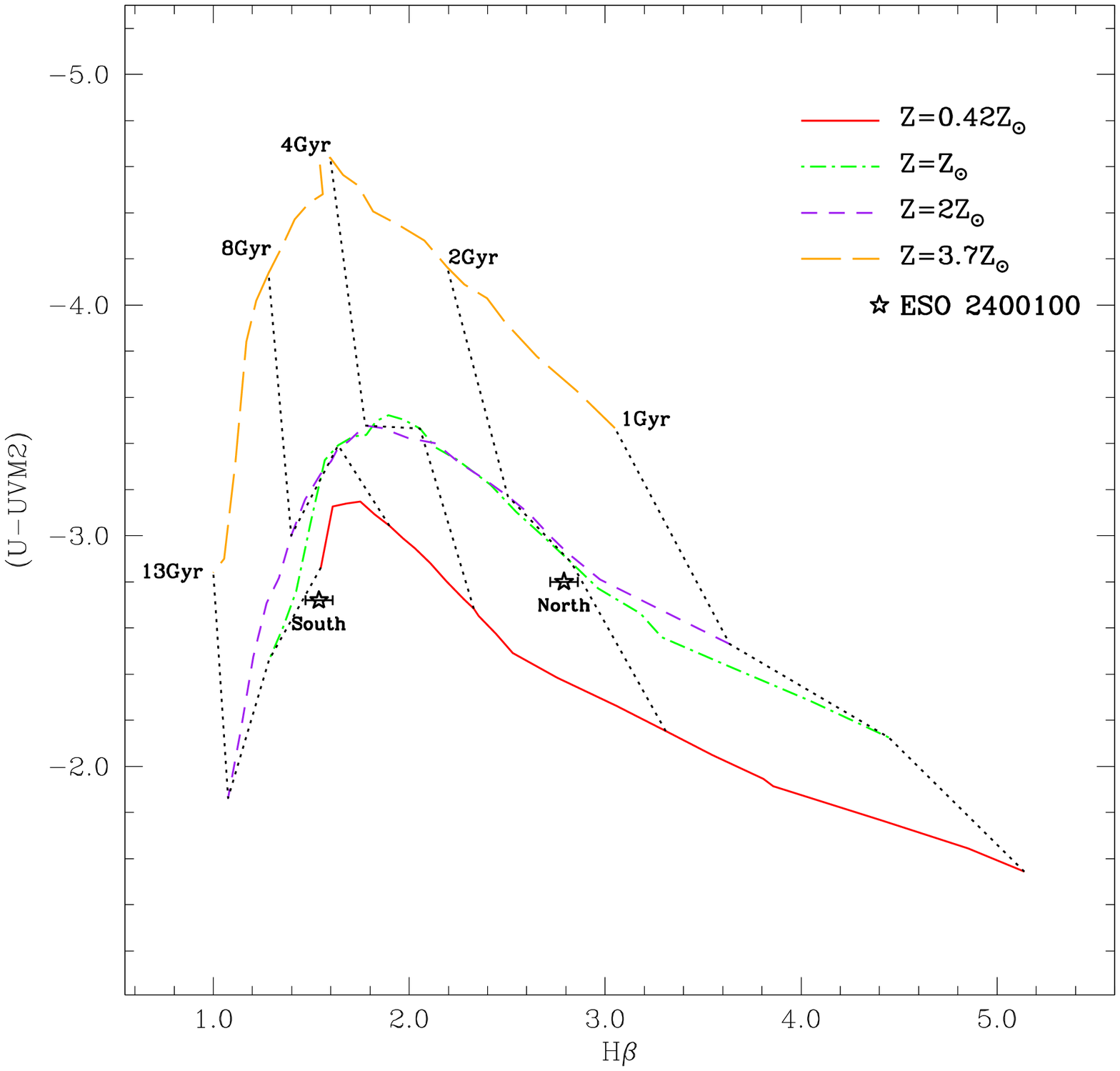}
     \includegraphics[width=7.5cm]{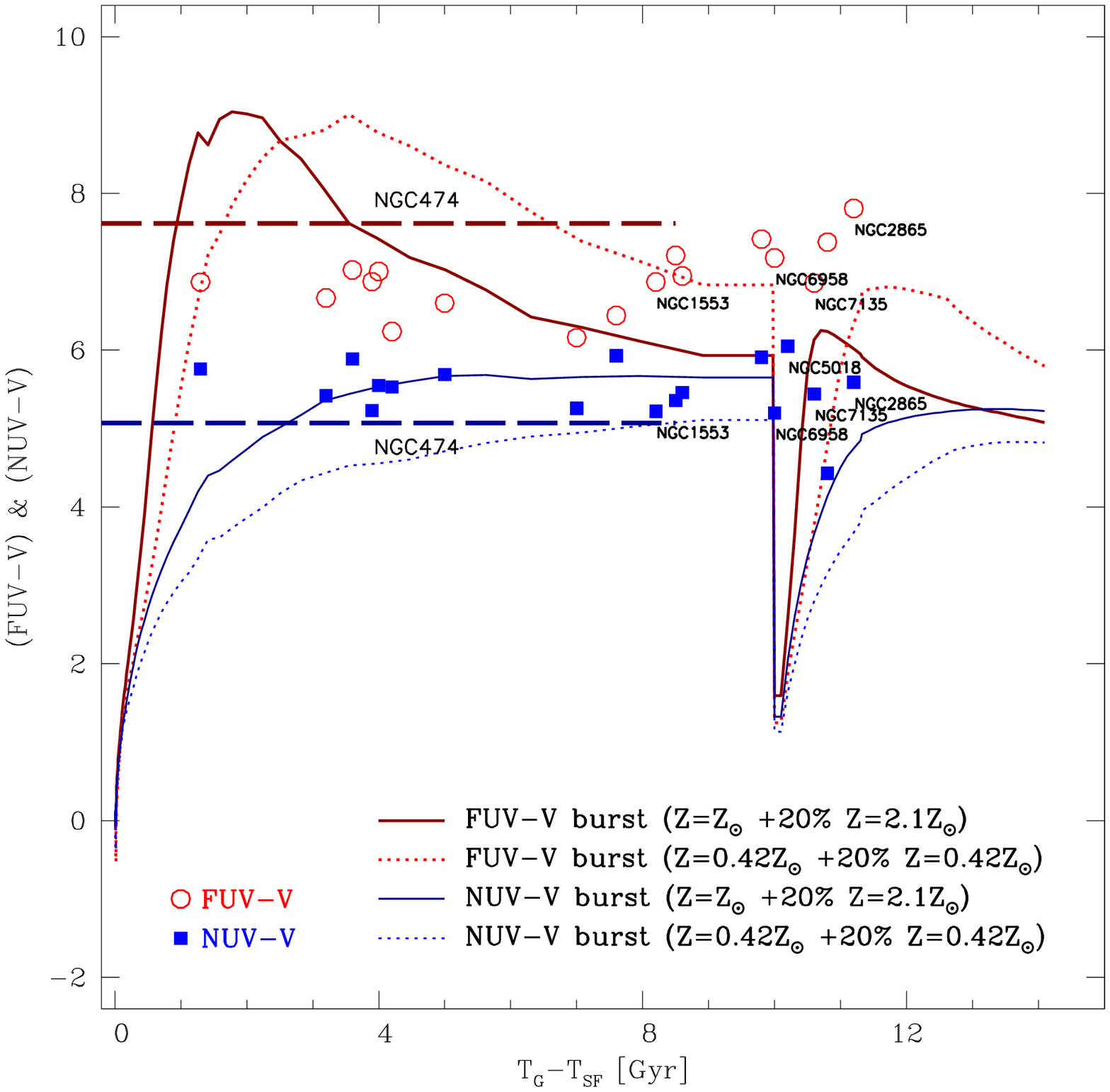}
     \includegraphics[width=7.5cm]{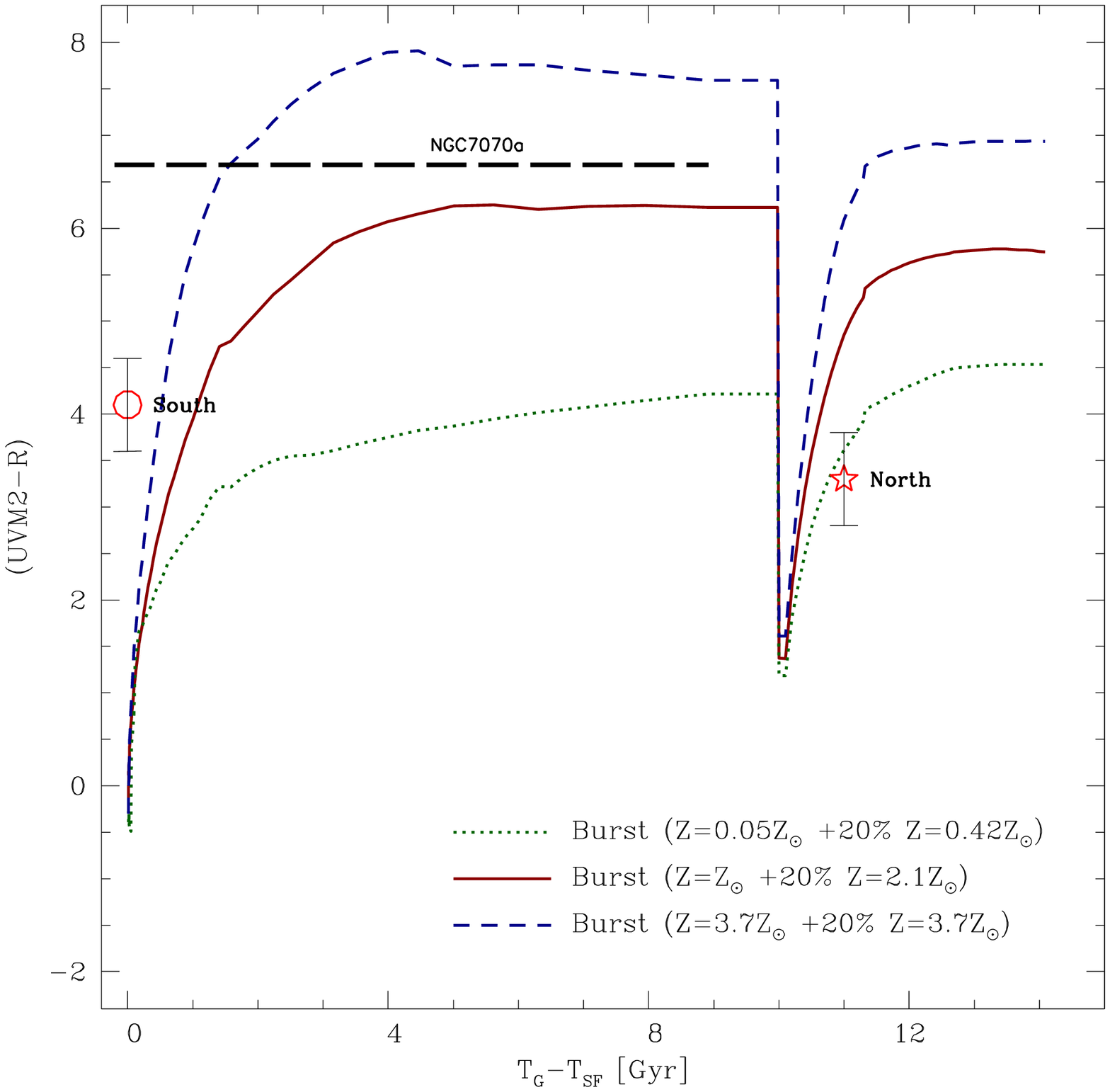}
     }
\caption{  Comparison of the theoretical relations, for different metallicities, 
and observations. ({\it Left}) (U-UVM2) -- H$\beta$ plane for ESO~2400100 
$a$ (South) and $b$ (North) nuclei.
We consider four different metallicities in the
range $Z$ 0.008 -- 0.070  with $Z_{\odot}$=0.019.
({\it middle}): Models for the evolution of the {\it GALEX} (UV - V) colors vs. age, 
expressed as the time since last SF episode, $T_G-T_{SF}$=13-$T_{SF}$. 
Two metallicities for the young and for the old stellar component 
($Z$=0.008/0.019 and $Z$=0.040/0.008 respectively), 
with a burst intensity fixed at 20\%, are assumed (see labels in the figure). 
No age can be assigned to NGC 474, so we plot it as a line at its {\it GALEX} (NUV - V) 
and (FUV - V) colors. Values for the  sample of 18 early-type galaxies analyzed by
\citet{Rampazzo07} are also plotted for comparison, with  shell galaxies labeled. 
({\it right}) {\it XMM-OM} (UV - R) colours of  NGC~7070A and ESO~2400100.  
The plotted models are for three different values of the young component 
($Z$=0.008/0.040/0.070) and old component
($Z$=0.001/0.019/0.070). Notice that the color 
of NGC~7070A is consistent with a metallicity higher than solar.
}
 \label{fig10}
   \end{figure*}

Figure~\ref{fig10} compares the theoretical predictions
from different SSP models  with  metallicities in the range
0.008$\leq$Z$\leq$0.070 with the observations.  Different  optical-UV colors
are plotted as a function of ``age" estimated as the time elapsed from the last episode of star formation
activity computed in different ways, depending on the available information. 
The left panel shows the nuclear line--strength index  H$\beta$  vs 
(U-UVM2) for the two nuclei of \eso. 
The H$\beta$ values are 
from \citet{Longhetti00} while the (U-UVM2)  (and  (UVM2-R), see later) nuclear
colors are calculated from the present photometry,
within an aperture of about 3\arcsec\
diameter centered on each nucleus, separated by  about
5\arcsec. 
The comparison with the models suggests an age of $\sim 2$ and  $\sim 13$
Gyr for the northern and southern nuclei, respectively.  This is consistent with the photometric
results of Fig.~\ref{fig9}, where the north, younger nucleus is significantly bluer, which is
likely to be the result of a contribution from a young stellar
population.

The estimate of 2 Gyr for the northern nucleus, as the age of the last
star formation event, should be considered an upper limit. In fact,
\citet{Longhetti00} did not correct the values of H$\beta$ for line
emission infilling. \citet{Rampazzo03} showed the presence of ionized
gas, whose filamentary distribution covers the two nuclei. A correction
for line infilling would deepen the H$\beta$ line,  increasing the value
of the line-strength  index, and would lead to a smaller age estimate
($<$ 2 Gyr).    

In the middle and right panels of Figure~\ref{fig10} we model the
colour evolution of an early-type galaxy which has had a recent 
burst of star formation. The burst
occurs at the age $T_G \sim$  10 Gyr. The intensity of the burst is 20\% and two
possible combinations for the metallicity for the old and young components
are shown. Of course, other values of the age at which the burst
occurs, its intensity, and the chemical composition could be
adopted. In the middle panel, we plot the {\it GALEX} UV - V
colors and the age of  the sample studied by \citet{Rampazzo07}. 
The age is plotted as  $T_G-T_{SF}$,
where $T_G$ is the canonical age of 13 Gyr, and $ T_{SF}$ is
derived from the $H\beta$ index.   In other words,
all galaxies are supposed to have initiated their star formation history
13 Gyr ago. 

The aim of this plot is to compare the observed properties of the galaxies
with a typical star formation history of a galaxy which
could have had an accretion episode which triggered a secondary
star formation episode. 

Since we do not have an age estimate for NGC 474,  we indicate the
(FUV-V) and (NUV-V) values for this galaxy with a dashed line. 
The galaxy's
colors are entirely consistent with those of other ``normal" and shell
galaxies. Without other elements, the color analysis suggests that
NGC~474 could be explained equally well either as the result of a recent
dry merger \citep[it has neither cold nor warm gas associated, see
e.g][]{Rampazzo06}, or as an old accretion event, which is recorded in
the shell system.   Its (FUV-V)  color is at the upper bound of
the FUV-V colors observed for shell galaxies in the \citet{Rampazzo07}
sample: we therefore suggest an age $T_G-T_{SF} \approx 10-12$ Gyr, 
in analogy with the age
range inferred from the  line--strength indices for the shell galaxies
in the above sample, which could be considered an upper limit to the true
age of NGC 474. 

In analogy with the previous panel  we model the evolution of a galaxy
in the (UVM2-R) vs. (T$_G$-T$_{SF}$) plane (Figure~\ref{fig10} right),
and we compare the results with the NGC 7070A and \eso\ data.  The 
(UVM2-R) color, 4.1$\pm$0.5 and 3.3$\pm$0.5 for the southern and the northern
nucleus of \eso, respectively, has a large uncertainty due to the limited
resolution ($\approx$ 2\arcsec) of the R-band images.  The positions of
the northern nucleus in the plot is consistent with a recent  merging
event that has rejuvenated its stellar population, unlike the southern
one, which appears composed of an older stellar population. 
We  do not have an independent estimate of the age of NGC
7070A, as for \eso, and unfortunately we cannot use the color to infer
its age,  since at about  $T_G-T_{SF}$=3 the (UVM2-R) color saturates,
and we can only constrain to be $T_G-T_{SF}>$3.

Given that no line-strength indices have been measured for NGC 7070A and
NGC 474, and the comparison with the models does not provide a reliable
age estimate, in the following section we use $T_{SF}$ = 2.4$\pm$0.6
Gyr for them, inferred from the average of the age  estimates for
the shell galaxies in the \citet{Rampazzo07} sample.

The  discovery of a double nucleus in  ESO~2400100 makes it a 
relatively peculiar and rare system  among shell galaxies. 
\citet{Forbes94} searched  for double nuclei in 29 shell galaxies,  
and 20\% of them with a
``possible" secondary nucleus undergoing slow disruption. Assuming a shell lifetime of 1 Gyr \citep[see e.g.][]{Hernquist87a,Nulsen89}
they estimate that the average lifetime of secondary nuclei is $<$200 Myr.
The  percentage provided by \citet{Forbes94} 
is probably an upper limit - as they state. Recent high resolution
HST analysis of shell galaxies  \citep{Sikkema07} did not evidence 
the presence of double nuclei.  In addition, both accretion models 
\citep[see e.g.]{Dupraz86,Kojima97}
and observations \citep[see e.g.][]{Longhetti00} indicate 
that shells are long lasting structures, i.e. the average shell lifetime  is probably 
longer, on average, than 1 Gyr. The estimate of the average 
time before coalescence of the two nuclei is then very uncertain. 
However, the lower fraction of double nuclei shell galaxies, 
with respect to the \citet{Forbes94} estimate, could be compensated  by
the longer shell life time in providing an average lifetime of 
a secondary nuclei not very different from the above value of   $<$200 Myr. 
 
As an alternative interpretation, the secondary nucleus 
in ESO~2400100 could be considered a chance superposition, 
e.g. with a group member \citep[see e.g.][]{Forbes94}. 
However, the kinematical analysis of \citet{Longhetti98a} and 
\citet{Rampazzo03} suggests that the two nuclei are strongly interacting, dismissing this alternative.  
The star formation history  of the two interacting nuclei, as deduced from Figure~\ref{fig10}, can be compared with the simulations for shell formation by
\citet{Kojima97}, which provide some predictions for the star
formation in the smaller accreted satellite and for the time of formation 
and duration of the induced shells.  In a radial merger, the star formation 
activity in the gas-rich satellite that was active before merging 
is completely truncated after its first passage through the primary member. 
The material that the primary member accretes from the satellite would produce
regular shells several 10$^8$ year after the truncation of the 
star formation activity.  In a
prograde merger instead the star formation is triggered, but shells and loops
start to develop only after the starburst phase is over.  
The duration of the shells is of the order of or slightly greater than 1 Gyr. 
The difference between different cases lies in the gas dynamics: gas
could be pushed at large radii by the interaction decreasing the gas
density or conveyed toward the center of the sinking satellite due to
the  deformation of the potential.  Line-strength indices
and UV observations suggest that the northern nucleus has had a very recent
star formation episode. Therefore, in the framework provided by  the \citet{Kojima97} simulations, we may be taking a snap-shot of an on-going accretion and shell formation event.

\subsection{Does the X-ray luminosity evolve with time after a merging episode?}

Our galaxies are located in low density environments, although they are
all in poor galaxy structures:  in particular,
NGC~474 is member of a poor, probably evolving group
\citep{Rampazzo06}. Although at different
levels, all are interacting systems, and as suggested by the previous
section, relatively ``young", as measured by  $T_{SF}$.

Age could be related to the X-ray emission in early-type galaxies, as
suggested by \citet{Sansom00} and \citet{Osul01}: early-type galaxies
showing fine structure -- like shells, dust-lane, etc. -- or a younger
spectroscopic age tend to have a smaller gaseous component, although the
dispersion is very large.

   \begin{figure*}
      \resizebox{16cm}{!}{
   \includegraphics[width=8.8cm]{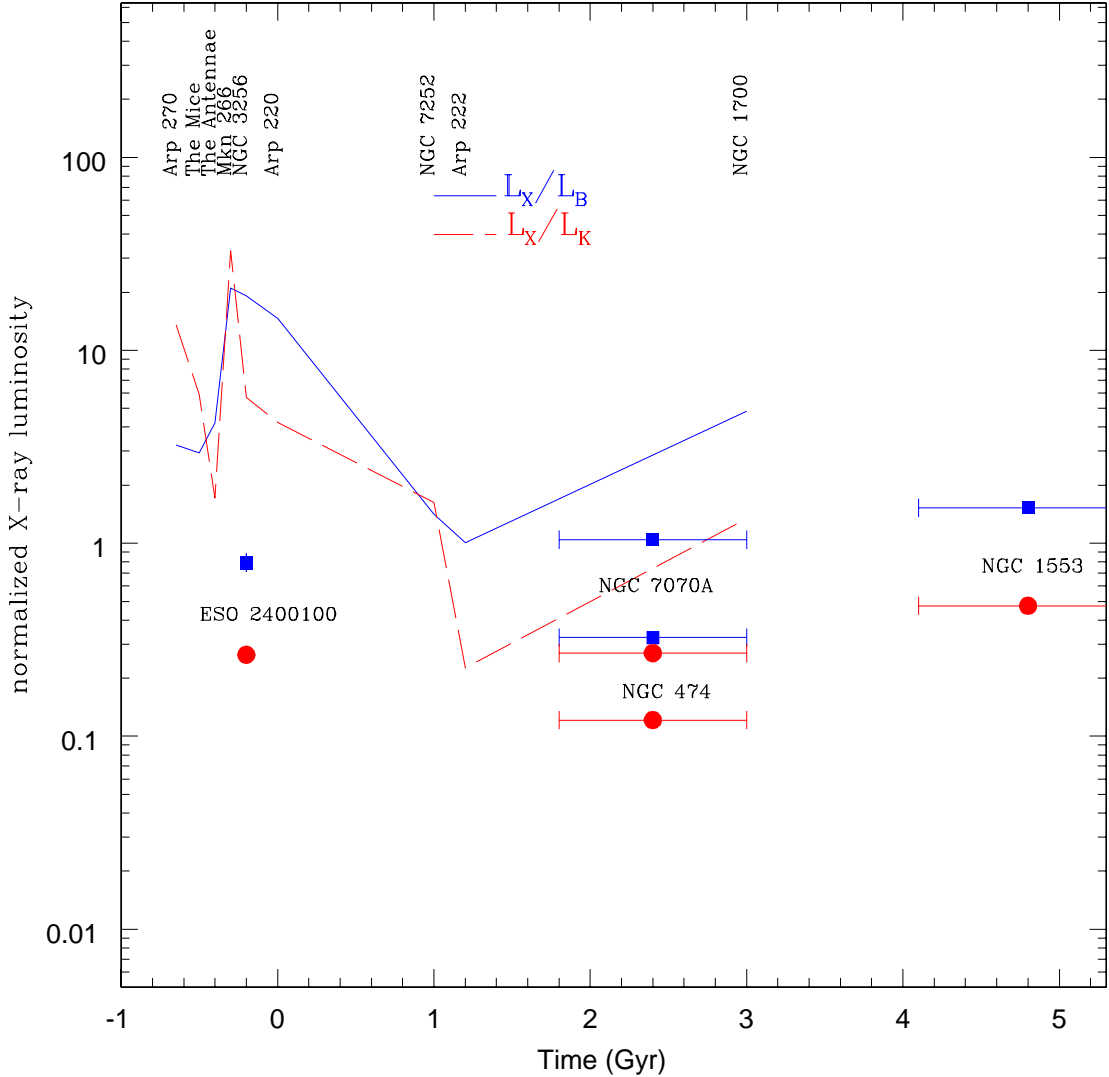}
}
   \caption{The evolution of X-ray luminosity in major-merging galaxies.
   The figure is adapted from \citet[][see their figure
   16]{Brassington07}. The solid and long-dashed lines represent the
   trend of  the L$_X$/L$_B$ and L$_X$/L$_K$ ratios measured by these
   Authors. The X-ray luminosity, in the 0.3-6.0 keV range,  is
   normalized to the B-band and the K-band luminosity. All luminosity
   ratios  are finally normalized to the proto-typical spiral galaxy NGC
   2403 as in \citet{Brassington07}. The $x-axis$ is an estimate of the
   merger age, where 0 age is defined to be the time of the nuclear
   coalescence. L$_X$/L$_B$ and L$_X$/L$_K$ of our shell galaxies are
   represented by (blue) full squares and (red) full circles
   respectively. The values of NGC~7070A plotted represent an upper
   limit of the contribution of the galactic emission, since overall 
   the emission is dominated by the presence of a moderate luminosity AGN
 (derived as explained in the text).}
 \label{fig11}
   \end{figure*}

All three shell galaxies have a relatively low gas content and low
overall luminosity: the only evidence of a hot ISM in \eso, at a level
of L$_X \sim 5 \times 10^{39}$ \ergs. NGC 474 is a relatively extreme
case of low luminosity object devoid of gas \citep{Rampazzo06}. 
NGC~7070A is dominated by a low luminosity nuclear component, and we
could only assign a rather low limit to any plasma component. A few
other examples of ``young" and X-ray faint objects are available in the
literature \citep[e.g.][]{fabbianoSchweizer}. Recently \citet{Osul04}
detected 3 underluminous elliptical galaxies, NGC~3585, NGC~4494 and
NGC~5322, all of which show evidence of recent dynamical disturbances,
including kinematically distinct cores as in the case of NGC~474
\citep{Hau96}.

We can analyze the X-ray luminosity in shell galaxies as a function of
their ``age", in analogy to what  \citet{Brassington07} have done in
their study of the evolution of the X-ray emission during a merging
between {\it gas rich, disk} galaxies and compare then with the
interaction/merging simulations developed by \citet{Cox06} and 
\citet{dercole00}. 

From their  analysis of nine merging systems believed to represent different 
phases of a major merging process, \citet{Brassington07} suggest that (1) 
the X-ray luminosity peaks $\approx$300 Myr before nuclear coalescence; (2) 
 $\approx$1 Gyr after coalescence the merger remnants are fainter compared to 
mature ellipticals while (3)  at a larger dynamical age ($\geq$3 Gyr) 
remnants start to resemble typical ellipticals in their hot gas content. 
Based on this, the above authors agree with the idea of an 
{\it halo regeneration} with time.  Can we extend this to
early type galaxies, where minor merger and/or accretion
events between gas-poor and smaller gas-rich
satellites take place?  Shell galaxies are a suitable set
to be compared with the evolutionary picture 
discussed by   \citet{Brassington07}:  the shell systems presented
above are very homogeneous, their luminosity profiles
and dynamics suggest the presence of a disk, and the shells suggest that a
recent accretion event has perturbed their potential.  In addition,
the rare case of the double nucleus in 
ESO~2400100 allows us to follow the X-ray luminosity
evolution before the companion has been digested.  We add another well
studied shell system, NGC 1553, observed with {\it Chandra} by \citet{Blanton01}. 
NGC 1553 has optical properties in line with those of our sample: it is an
S0 in a small group  interacting with an elliptical companion, NGC 1549.
Adopting a distance of 13.4 Mpc \citep{Tully88} we compute a X-ray luminosity 
in the range 0.3-6 keV of  L$_X$=2.16$\times 10^{40}$ erg~s$^{-1}$. 

In Figure~\ref{fig11}, adapted from Fig. 16 in \citet{Brassington07},  
we plot the relation between the X-ray
emission and the time elapsed from the merging episode for shell galaxies. 
As discussed in the previous section, 
for  NGC~474 and NGC~7070A  $Time$ is the average 
age of the last star formation episode obtained  
from the comparison between models and our UV 
and optical data set. For ESO~2400100, 
we assume $Time$ = -200 Myr as the age of coalescence. i.e. 
the estimated average  lifetime of 
secondary nuclei made by \citet{Forbes94}.  For NGC 1553, 
the study of  line-strength indices by 
\citet{Annibali07} provides an age of 4.8$\pm$0.7 Gyr as the 
age of the last burst of star formation.

Clearly the plane is very poorly sampled, both in $Time$ and in number of
objects at any given time \citep[as in][]{Brassington07}, so any intrinsic
scatter cannot be accounted for.  However, all 
shell galaxies appear systematically fainter than
the gas rich systems, at very comparable levels (only a factor of a few in the
different quantities plotted). 
ESO~2400100 is 1-2 orders of magnitude weaker that NGC~3256 or Mkn 266 in the
pre-merging \citet{Brassington07} sequence, where a peak in the relation is observed. 
The same is true for  NGC~474 at the assumed age of 2-3 Gyr. 
The comparison with NGC 7070A is complicated by the presence of the nuclear
source, which prevents us to obtain a reasonable estimate of a galactic
component. In section 4.2 we have estimated an upper limit for the hot ISM 
of a few $\times 10^{38}$
\ergs.  To estimate an upper limit to the emission from the binary population,
we assume that their contribution would be at or lower than the unabsorbed power law component in the spectral fit (see section 4.2, Fig~\ref{fig4}).  This amounts to $\sim 5 \times 10^{39}$
\ergs, which is the value used in Fig.~\ref{fig11}.  With these values, NGC 7070A is comparable to NGC 474, at the same
age.
NGC 1553, the oldest system, is a factor of a few fainter than  NGC 2434,
the reference ``prototypical" elliptical galaxy chosen by \citet{Brassington07}. 
We cannot ascribe these differences to the galaxy mass, since all appear to have a similar central velocity dispersions 
\citep[see Tab.~1; $\sigma$=180$\pm$3.9 for  NGC~2434; note that
$\sigma$=180$\pm$20 km~s$^{-1}$ for NGC 1553, ][]{Rampazzo88, Beuing02}. 

All in all, it appears that any age effect on luminosity 
is significantly less evident in these systems compared to the
\citet{Brassington07} sequence.  This is most likely related to the intrinsic
properties of the two sets: in gas-rich systems, merging would trigger a
significantly larger star formation episode, with the production of active
binary systems and shock heating of the available gas.  In gas poor systems,
even the acquisition of a smaller gas rich object will have limited effects.
There is however an intriguing difference in the gas content at
different $Time$.  Gas is found in \eso, before coalescence, and in NGC
1553, at a later stage, although at lower levels than the ``prototypical"
mature elliptical NGC 2434.  This is in qualitative agreement
with a picture in which
several gigayears are required to refurbish a galaxy of a  hot gaseous
halo after a merging responsible for its depletion \citep{Sansom00,
Osul01}, although the difference with NGC 2434 still remains to be properly understood. 

\citet{Cox06} examine the X-ray emission produced from hot gas during collision and mergers of disk galaxies.  Models include the effects of radiative cooling,
star formation, supernova feedback, and accreting super-massive black holes (BH).  The gas emission predicted in either of their models, i.e. ``with'' and ``without'' the central accreting Black Hole, is significantly higher than observed in either NGC~474 or NGC~7070A \citep[cf. Fig. 3 and Fig. 5 in][]{Cox06}.  Moreover, we notice that the X-ray emission of NGC~7070A  is dominated by the AGN. The AGN contribution is not considered in  the \citet{Cox06} models since the central accreting 
BH in merger simulations remain obscured for the majority of their lifetime. 
On the contrary, ESO 2400100 appears to be
more luminous than expected if we consider it a pre-merger, at age $\sim 0$ (see section 5.1).

No gas is detected in NGC 7070A or NGC 474: in the former this could easily be
explained by the nuclear emission, that could be fueled by the ambient gas.
Numerical simulations indicate that, in major mergers, roughly
half the gas is quickly funneled to the centre,
some causing a burst of star formation and much settling in a high
surface density, mostly molecular central disc;
the other half is ejected to large distances but
remains bound to the merger remnant. This high
angular momentum diffuse gas eventually falls back and may settle
into a more extended, large-scale disc \citep{Barnes,Bournaud, Naab}.
Several theoretical works suggest that galaxies of moderate 
optical luminosity and mass are likely to
develop galactic outflows, capable of expelling most/all of the gas shed by evolving
stars,  even without resorting to the heating energy  from a central AGN \citep{Ciotti, David}.  Partial or global galactic outflows could then
also be invoked to explain the lack of a hot ISM in these galaxies.  At the present time, a galactic outflow is observed only in NGC 3379, a very well studied nearby early-type galaxy, for which very deep Chandra observations are available \citep{Trinchieri}.

Our X-ray study shows that NGC~7070A hosts a moderate AGN activity. 
Does  it fit in the X-ray luminosity evolutionary scheme? 
Among the galaxies considered by \citet{Brassington07}, 
 Mkn~266 has a faint AGN,
and  Arp~220 possibly hosts a heavily shrouded AGN. They are in  the
pre-merging sequence, a few Myr before 
the coalescence of the nuclei. NGC~7070A is already a merger remnant.

Recently, \citet{Canalizo07} found a well developed shell
structure in the elliptical host of the QSO MC2 1635+119.
To reproduce the elliptical host and its shells a simple 
N-body model gives an estimate 
between $\approx$30 Myr and 1.7 Gyr for the age of the merger under reasonable
assumptions. On the other hand, the spectrum of the host
galaxy is dominated by a population of an age of 1.4 Gyr, indicating 
a strong starburst episode that may have occurred at the time of the 
merger event.  Although we do not have an accurate estimate
from line-strength indices of the time at which the last episode 
of star formation in NGC~7070A has occurred, the average estimate 
of the last star formation episode adopted  from the
comparison of our UV data with models is consistent
within the errors with the above estimate for QSO MC2 1635+119.

\section{Summary and conclusions}

We have studied the characteristics of  the X-ray emission of two shell systems, 
NGC~7070A and ESO~2400100 using XMM-Newton. We  also
analyzed their far UV emission using the XMM-Newton Optical
Monitor.  We include another shell galaxy NGC~474, for which we have
already presented the X-ray characteristics in \citet{Rampazzo06},
exploiting both original XMM-Newton OM images and {\it GALEX} archival data.

The XMM-Newton  spatial and spectral analysis  
suggest that a nuclear source is the dominant component in NGC 7070A. 
In ESO~2400100, the emission is due to a low temperature plasma
with a contribution from the collective emission of individual sources
in the galaxy. 

{\it GALEX} data of NGC~474 show that the extension of NUV emission 
is comparable with that of the optical image,  while the
FUV emission shows up only in the central regions of the galaxy.
Also  in the UVW1 and UVM2 filters  NGC~474, ESO~2400100
and NGC~7070A have extensions similar to that of the optical image.
XMM-OM UV images of ESO~2400100 show the presence of a
double nucleus. The shape of the luminosity profiles 
suggest that a disk component is present, 
confirming the morphological classification provided by {\tt NED} 
and the suggestion, from the kinematical study of
\citet{Sharples83}, that NGC 7070A is an S0 seen face-on.  This study
further suggests that the prominent dust lane is not yet in equilibrium,
indicating a recent accretion episode. 

We model line--strength indices and the far UV - optical colors
of the galaxies to infer the time elapsed from the last significant 
episode of star formation. 
From the comparison between  our galaxies
and shell galaxies in the \citet{Rampazzo07} sample we suggest that the (UV - optical) colors
of NGC~474 and NGC~7070A are consistent with a recent burst of star formation. 
We argue that the double nucleus in ESO~2400100 is indicative of an ongoing accretion event.
The combined UV and line-strength indices analysis suggests indeed
a very recent star formation episode in the northern nucleus 
\citep[see also simulations by ][]{Kojima97}.

Using the above estimates of the time elapsed from the last significant
episode of star formation we investigate whether the evolutionary scheme
 for gas-rich systems that would lead to mature ellipticals proposed by
\citet{Brassington07} could be applied to the same early-type galaxies
undergoing merging episodes.  

The time range is very poorly sampled, with only 4 objects, but  we span
almost the full range, since  \eso\ should be at $Time \sim -200$ Myr,
i.e.  before coalescence. We notice that shell galaxies are
systematically underluminous relative to gas-rich systems at similar
evolutionary stages, and that the $L_x/L_B$ or $L_x/L_K$ values vary
only a factor of a few, compared to factors of 100. The only indication
of a possible difference is in the gas content, which is present either
before or several Gyr after coalescence. This is consistent with a
picture in which several gigayears are required to refurbish a galaxy of
a  hot gaseous halos after a merging responsible for its depletion
\citep{Sansom00, Osul01}.

To fully understand whether  shell galaxies are the precursors of  relaxed
ellipticals, their more mature counterparts according to hierarchically
evolutionary scenarios, we need  to better define the  dynamical
and photometric time-scales of  the accretion/merging event and understand
whether there is an evolution of their 
X-ray properties linked to their ``age" based on a larger and better studied
sample of objects. The study of the UV emission in
early-type galaxies in connection with optical line-strength indices
\citep[see e.g.][]{Rampazzo07} could provide useful insight for timing
their photometric evolution.

\begin{acknowledgements}
We acknowledge partial financial support of the Agenzia Spaziale
Italiana under contract ASI-INAF contract
I/023/05/0. RG and RR acknowledge the support and kind hospitality
of INAF-Osservatorio Astronomico di Padova and Institut f\" ur Astronomie 
der Universit\" at der Wien, respectively.  We thank
Dr. Nora Loiseau of XMM-Newton User Support Group for her skillful assistance. 
{\it GALEX} is a NASA Small Explorer, launched in April
2003. {\it GALEX} is operated for NASA by California Institute of
Technology under NASA contract  NAS-98034. This research has made
use of the SAOImage DS9, developed by Smithsonian Astrophysical 
Observatory and of the NASA/IPAC Extragalactic Database (NED) which 
is operated by the Jet Propulsion Laboratory, California Institute of
Technology, under contract with the National Aeronautics and Space
Administration. {\tt IRAF} is distributed
by the National Optical Astronomy Observatories, which are operated
by the Association of Universities for Research in Astronomy, Inc.,
under cooperative agreement with the National Science Foundation.
The Digitized Sky Survey (DSS) was produced at the
Space Telescope Science Institute under U.S. Government grant NAG
W-2166. The images of these surveys are based on photographic data
obtained using the Oschin Schmidt Telescope at the Palomar
Observatory and the UK Schmidt Telescope. The plates were processed
into the present compressed digital form with the permission of
these institutions.
\end{acknowledgements}

\end{document}